\documentclass[a4paper,11pt]{article}
\usepackage{jheppub} 
\usepackage{lineno}

\usepackage{amsmath}
\usepackage{amssymb}
\usepackage{amsfonts}
\usepackage{amsthm}
\usepackage{amsbsy}
\usepackage{array}
\usepackage{mathrsfs}
\usepackage{physics}
\usepackage{mathtools}
\usepackage{hyperref}
\usepackage{subcaption}
\usepackage{dsdshorthand}

\usepackage{tikz}
\usetikzlibrary{decorations.markings}

\usepackage{tikz}
\usetikzlibrary{decorations.pathmorphing, arrows.meta}
\usetikzlibrary{shapes.geometric}
\title{\boldmath Memory detectors and fully constraining Faddeev-Kulish dressings in QED and gravity}

\author[a]{Brett Oertel,}

\affiliation[a]{Department of Physics, Yale University,\\217 Prospect St, New Haven, CT 06511, USA}

\emailAdd{brett.oertel@yale.edu}

\abstract{We show that for both QED and perturbative quantum gravity, time dependent Faddeev-Kulish dressings can be fully constrained either by symmetry, or by demanding that they reproduce the correct outgoing classical memory contribution. For gravity, we show that using this dressing to construct finite-time Fock spaces allows us to recover both the first order gravitational memory, as well as higher order Christodoulou contributions from the  gravitational field. We explain how these higher order perturbative corrections arise in inclusive in-in calculations. }

\begin{document}
\definecolor{yaleblue}{rgb}{0.06, 0.3, 0.57}
\newcommand\BOnote[1]{{\color{yaleblue}{\bf BO: #1}}}
\maketitle
\flushbottom

\section{Introduction} 
\subsection{A very brief introduction to the memory effect}
\label{sec:intro}
The memory effect in electromagnetism or gravity can be loosely thought of as the angular components of the field near future null infinity asymptoting to different values at early and late retarded times. Thus, it provides an interesting physical measurement which can be performed at far distances from a scattering process. The full non-linear memory effect was first calculated in classical gravity by Christodoulou \cite{Christodoulou:1991cr}. It has also been calculated in perturbation theory \cite{Braginsky:1987kwo, Thorne:1992sdb, Wiseman:1991ss}. More recently, Bieri and Garfinkle used the Weyl tensor to rederive the memory effect in gravity to first order in perturbation theory in a gauge invariant manner \cite{Bieri:2013ada}. \footnote{They also calculated it in electromagnetism \cite{Bieri:2013hqa}. } 

Much interesting work has also been done recently to formalise how to calculate the memory effect in QED and perturbative quantum gravity using QFT scattering technology \cite{Gabai:2016kuf, Pasterski:2015zua, Strominger:2014pwa, Choi:2017bna, Prabhu:2022zcr, Hirai:2019gio, Hirai:2020kzx, Hirai:2022yqw, Moult:2025njc} and these methods have recently been generalised to include scattering processes containing external hard gravitons using detectors and time dependent, gauge fixed Faddeev-Kulish (FK) dressed Fock spaces \cite{Oertel:2026wsm}. It is important to note that by extending these results to hard gravitons, the authors of \cite{Oertel:2026wsm} manage to capture the contributions to memory from gravitational waves emitted in scattering, sometimes referred to as the `non-linear' Christodoulou memory (this contribution is neglected in treatments at first order in perturbation theory, for example in \cite{Bieri:2013ada}). \footnote{The non-linear memory was also recently calculated in scattering theory without using dressings in \cite{Georgoudis:2025vkk}.}

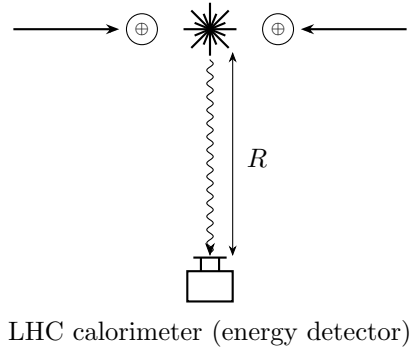
\begin{figure}[h!]
    \centering
\begin{tikzpicture}[
    every node/.style={font=\small},
    wavy/.style={decorate, decoration={snake, amplitude=1.2pt, segment length=5pt}},
]
 
\draw[-{Stealth[length=5pt]}, thick] (-2.6, 0) -- (-1.2, 0);
\node[circle, draw, inner sep=1.5pt] at (-0.9, 0) {\tiny $\oplus$};
\draw[-{Stealth[length=5pt]}, thick] (2.6, 0) -- (1.2, 0);
\node[circle, draw, inner sep=1.5pt] at (0.9, 0) {\tiny $\oplus$};
 
\foreach \a in {0,45,...,315} { \draw[thick] (0,0) -- ++(\a:0.35); }
\foreach \a in {22.5,67.5,...,337.5} { \draw[thick] (0,0) -- ++(\a:0.22); }
 
 
\draw[wavy, -{Stealth[length=5pt]}] (0, -0.4) -- (0, -3.0);
 
\node[right=10pt] at (0, -1.7) {$R$};

\draw[{Stealth[length=4pt]}-{Stealth[length=4pt]}] (0.3, -0.3) -- (0.3, -3);

\draw[thick] (-0.22, -3.02) -- (0.22, -3.02);
\draw[thick] (-0.12, -3.02) -- (-0.12, -3.2);
\draw[thick] (0.12, -3.02) -- (0.12, -3.2);
\draw[thick] (-0.3, -3.2) rectangle (0.3, -3.6);
 
\node[below=3pt] at (0, -3.6) {LHC calorimeter (energy detector)};
 
\end{tikzpicture}
    \caption{The diagram shows protons colliding at the LHC.  A calorimeter measures the energy of particles produced in the process. The calorimeter sits at a distance $R$ which is much larger than the distance scale of the scattering process. Thus, we may describe the calorimeter using an energy detector at future null infinity, and the outgoing particle state using asymptotic Fock spaces defined at $t=\infty$.}
    \label{fig:energydetector}
\end{figure}

That a description of memory was achieved using detector technology should not be surprising since detectors formalise measurements made at future null infinity. \footnote{For more on detectors in perturbative field theories see \cite{Christ:1972ms, Brandt:1970kg, Sterman:1975xv, Basham:1978zq, Hofman:2008ar, Caron-Huot:2022eqs, Moult:2025nhu, Herrmann:2024yai}. } For example, energy detectors measure the energy released in a particle scattering experiment, as shown in Figure \ref{fig:energydetector}.

However, things can be more subtle when considering memory detectors \cite{Oertel:2026wsm}. This is partly because the memory effect is encoded into the asymptotic Fock spaces using Faddeev-Kulish (FK) dressings. In fact we will show here that demanding this fully constrains the gauge fixed form of these dressings. Furthermore, one may wish to take time dependence into account. Strictly speaking, FK dressings have $t$ dependence \cite{Kulish:1970ut, Ware:2013zja}, but this $t$ dependence is often dropped. However, it seems likely that if there is any hope to properly describe the asymptotic Fock space using FK dressings, it would lie in a time-dependent description. Indeed, IR divergences in theories with long range forces are intimately related to the late time behaviour of particles (for a recent detailed discussion on this, see the excellent paper by Lippstreu \cite{Lippstreu:2025jit}). \footnote{That a $t$ dependent formalism is necessary is also noted in \cite{Hirai:2022yqw}.}  Here we will show that in order to obtain mathematically consistent results without simply dropping the $t$ dependence of the FK dressings one must carefully consider the order of limits in inclusive memory calculations. 

As a first foray into the limits involved it is useful to think about a real memory experiment. In gravity, the memory effect measured after a scattering event is given to first order in perturbation theory by \cite{Bieri:2013ada}
\begin{equation}
   D^AD^B\Delta^{\mathrm{GR}}_{AB}=-E^{(3)}_{rr}(\infty)+E^{(3)}_{rr}(-\infty)+8\pi G F \ ,
\end{equation}
where $E^{(3)}_{rr}(u)\sim 1/r^3$ is the electric part of the Weyl tensor to leading order in a large $r$ expansion at constant retarded time $u$, and $F$ is the energy per unit solid angle radiated to infinity due to massless particles, which to first order in perturbation theory does not include gravitons. In \cite{Oertel:2026wsm} it is shown that acting the memory detector on states including external hard gravitons leads to the same formula, but with $F$ including energy contributions from gravitons produced in the scattering event. This was already predicted by Thorne in 1992 \cite{Thorne:1992sdb}. In the scattering technology developed here and in \cite{Oertel:2026wsm} the hard graviton contributions are higher order in perturbation theory since attaching outgoing gravitons to Feynman diagrams requires including additional vertices. Thus, as we will argue in this paper, inclusive in-in calculations using the memory detector defined in \cite{Oertel:2026wsm} in fact capture higher order perturbative corrections through higher order Feynman diagrams.

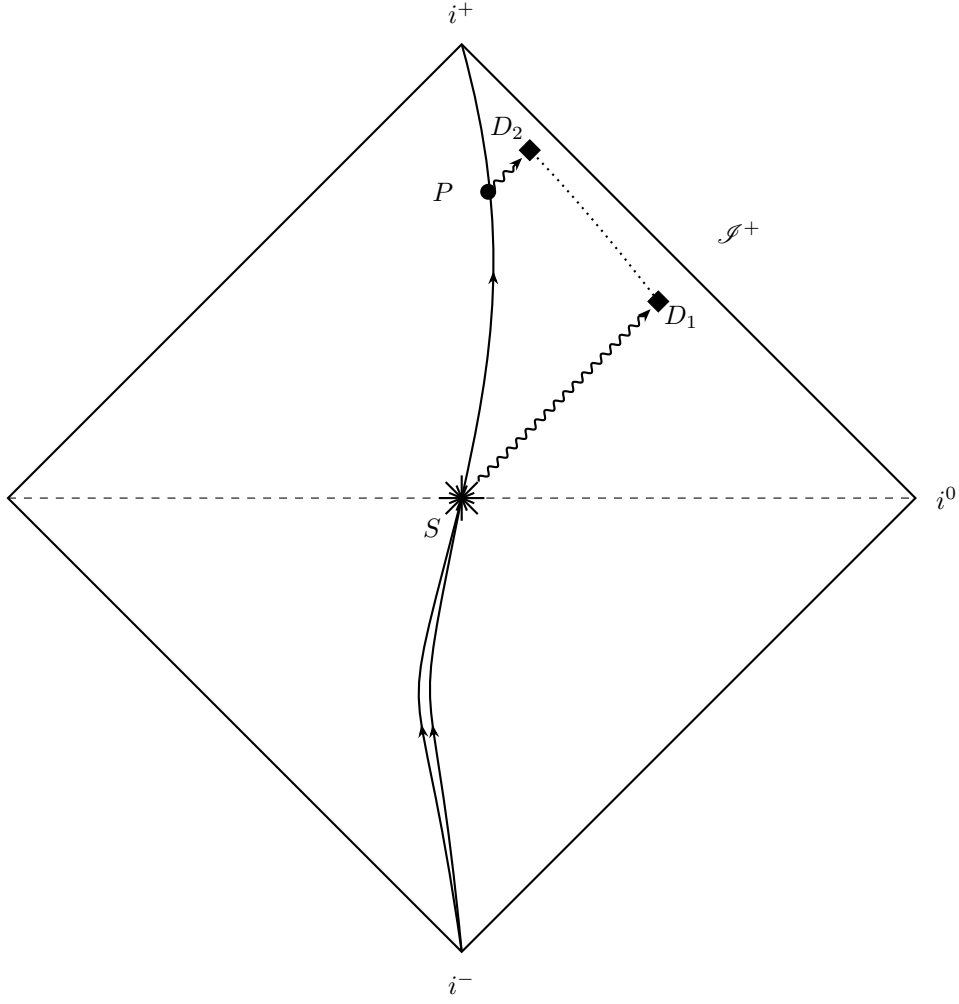
\begin{figure}[h!]
\centering
\begin{tikzpicture}[scale =1,
    every node/.style={font=\small},
    wavy/.style={decorate, decoration={snake, amplitude=1.2pt, segment length=5pt}},
    wavyarrow/.style={decoration={snake, amplitude=1.2pt, segment length=5pt,
        post length=3pt}, decorate, -{Stealth[length=5pt]}, thick},
    midarrow/.style={thick, postaction={decorate, decoration={markings,
        mark=at position 0.5 with {\arrow{Stealth[length=5pt]}}}}},
]

\coordinate (iplus) at (0, 6);
\coordinate (iminus) at (0, -6);
\coordinate (i0) at (6, 0);
\coordinate (ileft) at (-6, 0);

\draw[thick] (iplus) -- (i0) -- (iminus) -- (ileft) -- cycle;

\node[above=4pt] at (iplus) {$i^+$};
\node[below=4pt] at (iminus) {$i^-$};
\node[right=4pt] at (i0) {$i^0$};
\node[above right=4pt] at (3.1, 3.1) {$\mathscr{I}^+$};

\draw[dashed] (ileft) -- (i0);

\coordinate (S) at (0, 0);
\foreach \a in {0,45,...,315} { \draw[thick] (S) -- ++(\a:0.3); }
\foreach \a in {22.5,67.5,...,337.5} { \draw[thick] (S) -- ++(\a:0.18); }
\node[below left=4pt] at (S) {$S$};

\draw[midarrow] (0, -6)
    .. controls (-0.4, -1.8) and (-0.7, -3.5) .. (0, 0);
\draw[midarrow] (0, -6)
    .. controls (-0.6, -1.8) and (-0.9, -3.5) .. (0, 0);

\draw[thick, midarrow] (0, 0)
    .. controls (0.4, 1.8) and (0.7, 3.5) .. (0, 6);
\node[left=5pt] at (0.2, 4.05) {$P$};

\draw[thick, wavyarrow] (0.22, 0.22) -- (2.5, 2.5);

\coordinate (D) at (3.5, 1.5);

\draw[dotted, thick] (2.6, 2.6)
    .. controls (2.3, 3) and (1.5,4 ) ..  (0.9, 4.6);

\node[fill, draw, diamond, inner sep=2pt] at (2.6, 2.6) {};
\node at (2.9, 2.4) {$D_1$};
\node[fill, draw, diamond, inner sep=2pt] at (0.9, 4.6) {};
\node at (0.6, 4.9) {$D_2$};
\fill (0.35, 4.05) circle (3pt);

\draw[thick, wavyarrow] (0.3, 4) -- (0.8, 4.5);

\end{tikzpicture}
\caption{Penrose diagram showing two massive particles colliding at the origin $S=\{u=0,r=0\}$. They form a single massive outgoing particle which follows a timelike geodesic toward $i^+$, while the massless radiation produced propagates along null rays towards $\mathcal{I}^+$. A detector sits near $\mathcal{I}^+$ and travels along a timelike geodesic toward $i^+$, and receives two signals after $u=0$: the outgoing massless radiation which arrives at $D_1=\{u=0,r=R\}$ and the field response produced by the massive particle at $P$, which arrives at $D_2=\{u=\delta t, r=R\}$. Here $\delta t$ is large enough that the field measured is time-independent but small enough that it still satisfies $c \ \delta t\ll R$. The first signal contributes to $F$ (gravitons in the first signal contribute to the non-linear Christodoulou memory) whilst the second contributes to $E^{(3)}_{rr}(\infty)$.}
\label{fig:penrose}
\end{figure}

In a real experiment, $E^{(3)}_{rr}(\infty)$ is not measured at $u=\infty$ but rather at some $u$ large enough that the particles produced in the scattering event are approximately free. In fact, the situation is slightly more complicated. A real detector is not truly at future null infinity, but rather at some finite distance $R$ from the scattering event. When calculating $E^{(3)}_{rr}(\infty)$ one first takes the large distance limit $r\rightarrow\oo$ at constant $u$, neglecting terms like $u/r$, and then the limit $u\rightarrow\oo$. However, this approximation to a real detector is only valid so long as $E_{rr}^{(3)}(\infty)$ is measured at large enough $u$ that a steady state is reached (i.e. the outgoing particles are `free'), but where $u$ is still small enough that neglecting the $u/R$ terms is still valid. We depict this using a Penrose diagram in Figure \ref{fig:penrose}.

Thus we see that even in the classical case it is important to think about limits carefully when calculating memory.  In fact, we will see that the order of limits we just described in the classical theory will be exactly mirrored in the quantum theory! This is a reflection of the general philosophy of this paper which is that one can understand FK dressings better by understanding how to calculate memory. In the next two subsections we will discuss FK dressings in more detail. In particular, we will focus on what remains to be understood about them, and what advances we claim to  make in this paper.
\subsection{A very brief introduction to Faddeev-Kulish dressings}
\label{introFK}

FK dressings play a critical role in the memory calculations we will perform. However, FK dressings historically arose for a different reason, namely to resolve the IR divergences due to virtual photons and gravitons in the S-matrices of QED and perturbative quantum gravity. These IR divergences can be written in a very succinct way for QED and perturbative quantum gravity according to the Abelian exponentiation theorem \cite{Yennie:1961ad, Weinberg:1965nx}. Consider an amplitude $\cA$ for some scattering process, and let $\Lambda_{\mathrm{upper}}$ be an IR cut-off of the energy of all exchanged photons or gravitons and $\Lambda_{\mathrm{lower}}$ be another IR energy scale where $\Lambda_{\mathrm{lower}}<\Lambda_{\mathrm{upper}}$. Then, we can write

\begin{equation}
    \cA=\left(\frac{\Lambda_{\mathrm{upper}}}{\Lambda_{\mathrm{lower}}}\right)^{\a+i\b}\cA^0(\Lambda_{\mathrm{upper}}) \ .
\end{equation}
Here $\cA^0(\Lambda_{\mathrm{upper}})$ is the amplitude with exchanged photons and gravitons restricted to energies greater than $\Lambda_{\mathrm{upper}}$ and the IR divergences occur as we take $\Lambda_{\mathrm{lower}}\rightarrow 0$ which corresponds to including exchanged photons and gravitons of arbitrarily small energies in $\cA$. Note that $\a$ and $\b$ are real and depend on the kinematics of the scattering process. In 1970 Faddeev and Kulish \cite{Kulish:1970ut} (building on earlier work by Chung, Kibble, Dollard \cite{Chung:1965zza, Dollard:1964cok, 10.1216/RMJ-1971-1-1-5, Kibble:1968oug, Kibble:1968sfb}) constructed operators which dress external states in massive QED to cancel both the soft divergences which correspond to the real part $\a$ and the Coulomb phase divergence which corresponds to the $i\b$.  

The dressing operators which cancel the real divergences are what are commonly referred to as `Faddeev-Kulish' or `FK' dressings. They will form the main focus of this paper. The imaginary divergence is known as the Coulomb phase divergence and it is cancelled by `Dollard's operator'. This is also an important part of the story, for a recent discussion on this in QED see \cite{Lippstreu:2025jit}. In QED, FK dressings essentially dress outgoing charged particles with an infinite number of soft photons. FK dressings have also been constructed in gravity \cite{Ware:2013zja}, and they dress outgoing particles with infinitely many soft gravitons. 

Whilst FK dressings do cancel IR divergences due to soft, virtual photons or gravitons, they are far from a perfect fix. Before it can be claimed that Faddeev-Kulish dressings (including the imaginary part) can be used to construct an IR finite S-matrix there are many outstanding issues to resolve. Let us list these issues.
\begin{itemize}
    \item There are remaining non-integrable IR divergences which occur at exceptional momenta configurations such as forward scattering. Whilst these do not appear in sufficiently inclusive calculations, it would be nice to resolve them at the level of S-matrix elements.
    \item There are arbitrary choices of scales, such as $\Lambda_{\mathrm{upper}}$ which is encoded into the FK dressings, and a reference time $t_0$ which is encoded into Dollard's operator. As far as we know it remains to be shown that physical quantities do not depend on these.
    \item The S-matrix obtained using FK dressed states does not appear to satisfy a reasonable set of axioms when trying to describe the Fock space as a Hilbert space \cite{Prabhu:2022zcr}.
    \item There are gauge fixing functions $c_{\mu}$ and $c_{\mu\nu}$ which are not fully fixed in the standard constructions of FK dressings. Certain choices may lead to violation of the Steinmann relations \cite{Hannesdottir:2019opa}.
\end{itemize}
In this paper we only claim to make progress on the final issue listed (though we do not investigate the Steinmann relations). In the next subsection we will briefly discuss the main idea involved and its physical implications.

\subsection{Our claims and their implications}
We will discuss our claims in the context of perturbative quantum gravity but a similar story holds for QED in the obvious manner. In \cite{Ware:2013zja} FK dressings were constructed for perturbative quantum gravity and shown to depend on a gauge-fixing function $c_{\mu\nu}(k,p)$ where $k$ is a null momentum and $p$ is a null or time-like momentum.  The authors show that the function $c_{\mu\nu}$ must satisfy $c_{\mu\nu}=c_{\nu\mu}$ , $c_{\mu\nu}k^{\mu}=p_{\nu}$ , and $c^{\mu\nu}c_{\mu\nu}-(c^{\mu}{}_{\mu})^2/2=0$ in order to enforce gauge invariance and to not introduce additional singularities. The most general ansatz up to terms killed by physical polarization vectors is
\begin{equation}
    c_{\mu\nu}(k,p) = \frac{q_{\mu}p_{\nu}+q_{\nu}p_{\mu}}{q\cdot k}-\frac{k\cdot p}{(q\cdot k)^2}q_{\mu}q_{\nu} \ ,
\end{equation}
where  $q(k)$ is a null vector which is a function of $k$ with no fixed point. \textbf{The claim in this paper is that one must choose $q(k)=(1,-\hat{\bk})$ (up to a rescaling which drops out of $c_{\mu\nu}$) in order to reproduce the correct outgoing memory effect.} Note the following:
\begin{itemize}
\item This is the same as the choice made in \cite{Ware:2013zja, Choi:2017bna}. Here we will show that it is the unique choice which gives the correct memory due to the outgoing state as proven in Appendix \ref{uniquenessproof}. Note that equivariance motivates the choice of $q$ but need not be assumed. 
\item This corresponds to a non-vanishing BMS charge and ensures that the outgoing memory has no $\ell=0$  and $\ell=1$ parts.  \footnote{This may be a fix to the statement that the BMS charge vanishes in \cite{Choi:2017bna}, though it is likely the authors were simply referring to the $\ell\geq2$ parts.}
\item The gauge invariant object in the FK dressings is $\e^{\mu\nu}(k)(p_{\mu}p_{\nu}/k\cdot p-c_{\mu\nu})$. In general, different choices of $\e^{\mu\nu}(k)$ may or may not satisfy $\e^{\mu\nu}c_{\mu\nu}=0$. However, a gauge transformation which modifies $\e^{\mu\nu}c_{\mu\nu}$ will modify $\e^{\mu\nu}(k)p_{\mu}p_{\nu}/k\cdot p$ by a compensating amount. For example, one can show that the choice in \cite{Ware:2013zja} which leads to $\e^{\mu\nu}c_{\mu\nu}=0$ gives the same BMS charge as our choice which satisfies $\e^{\mu\nu}c_{\mu\nu}\neq0$.
\item Cancellation of IR divergences gives the constraint $c_{\mu\nu}(p,k)I^{\mu\nu\rho\sigma}c_{\rho\sigma}(p',k)=0$ for all $p,p'$ where $I^{\mu\nu\rho\sigma}=\eta^{\mu\rho}\eta^{\nu\sigma}+\eta^{\mu\sigma}\eta^{\nu\rho}-\eta^{\mu\nu}\eta^{\rho\sigma}$ \cite{Ware:2013zja, Choi:2017bna} which is quadratic in $c_{\mu\nu}$ and weaker than the constraint that the dressing reproduce the correct outgoing memory, which is linear in $c_{\mu\nu}$.
\end{itemize}
One may wonder why we would attempt to fix only one of the issues discussed in the Faddeev-Kulish dressings. The answer is that the author is interested in calculating memory correlators, and so began with the assumption that the classical memory should be derivable. The other issues in the FK dressings can be avoided by focusing on certain memory observables, for example in-in memory correlators at a certain order in perturbation theory where the sum over outgoing states yields an IR finite result. Thus, we may obtain useful physical results even without understanding our framework fully. This has always been the case in quantum field theory, and it is important to tread a line between rigour and practicality. In fact, our theoretical understanding can often be improved simply by trying to calculate physical observables, without worrying overly about details of the underlying theory.  Nevertheless, even within memory calculations the remaining issues are important if one wants to use FK dressings to calculate the contributions to loop order. For more on how our results relate to memory correlators see Section \ref{ininsection}.

\section{QED}
Here we will work with massive scalar QED. We require that our charged particles are massive in order to avoid problematic collinear divergences. On the other hand, we use a scalar field as opposed to fermions purely for convenience, and a similar story applies for the spin $1/2$ case. We will employ the field expansions in the `interaction picture'
\begin{equation}
\begin{aligned}
    &\phi(x) = \int_{p^0>0}\frac{d^4p}{(2\pi)^{3}}\delta(-p^2-m^2)(e^{ipx}a(p)+e^{-ipx}b^{\dagger}(p)) \ ,\\
    &A_{\mu}(x) = \int_{p^0>0}\frac{d^4p}{(2\pi)^{3}}\delta(-p^2)\sum^2_{r=1}(\a_r(p)\e^{r*}_{\mu}(p)e^{ipx}+\a_{r}^{\dagger}(p)\e^{r}_{\mu}(p)e^{-ipx}) \ ,
\end{aligned}
\end{equation}
where we are using the constant time commutation relations
\begin{equation}
    \begin{aligned}
        &[a(p),a^{\dagger}(q)] = 2\w_p (2\pi)^{3}\delta^{3}(\bp-\bq) \ ,\\
        &[b(p),b^{\dagger}(q)] = 2\w_p (2\pi)^{3}\delta^{3}(\bp-\bq) \ ,\\
        &[\a_r(p),\a_s^{\dagger}(q)] = 2\delta_{rs}\w_p  (2\pi)^{3}\delta^{3}(\bp-\bq) \ .
    \end{aligned}
\end{equation}
In canonical quantisation the canonical scattering Fock spaces are formed by acting on the vacuum with creation or annihilation operators defined on constant time slices at infinitely large $|t|$, i.e. \footnote{Here and throughout this paper we consider all operators as distribution valued operators, and limits are to be formally taken after integrating S-matrix products against test functions of momenta. However, we can only be so rigorous. In long range forces, the limit $t\rightarrow \oo$ does not exist, and whilst this is exactly what the FK dressing and Dollard's operator are trying to fix, there are remaining issues as we have discussed. For our purposes, we will assume that the eventual goal is to calculate in-in observables for which these subtleties do not matter. For more on this topic see \cite{Lippstreu:2025jit}.}
\begin{equation}
    \begin{aligned}
        &\bra{\mathrm{out}}=\lim_{t\rightarrow+\oo}\bra{0}a_{t}(p_1)\cdots b_{t}(p_{n'_s})\a_{t}(p_{n'_s+1})\cdots\a_{t}(p_{n'_s+n'_{\gamma}}) \ ,\\
        &\ket{\mathrm{in}}=\lim_{t\rightarrow-\oo}a^{\dagger}_{t}(k_1)\cdots b^{\dagger}_{t}(k_{n_s})\a^{\dagger}_{t}(k_{n_s+1})\cdots\a^{\dagger}_{t}(k_{n_s+n_{\gamma}})\ket{0} \ .
    \end{aligned}
\end{equation}
Here we have suppressed the polarization index on the photon creation and annihilation operators. The above are an out state with $n'_s$ massive, charged scalar particles and $n'_{\gamma}$ photons and an in state with $n_s$ massive, charged scalar particles and $n_{\gamma}$ photons. For ordinary scattering calculations this limit is often taken implicitly, and the ordinary scattering Feynman rules and LSZ formula are derived by assuming this limit exists. In fact, this is not true for theories with long range forces (like QED and gravity) and this leads to IR divergences like the Coulomb phase divergence. There are also soft divergences: if one forms S-matrix elements from an in and out state and takes the soft limit of an external photon, then the Weinberg soft theorem tells us that we encounter an IR divergence which scales as the inverse of the soft photon's energy \cite{Weinberg:1995mt}:
\begin{equation}
    \lim_{\w_{\gamma}\rightarrow 0}\bigg|\braket{\mathrm{out}}{\mathrm{in}}\bigg|\sim \lim_{\w_{\gamma}\rightarrow 0}\bigg|\frac{1}{\w_{\gamma}}\bigg| \sim \oo \ .
\end{equation}
Both these IR divergences can be cured by using FK dressed states, where each charged particle is dressed with an exponential factor. For example,  a dressed one particle in state consisting of a massive, charged particle with charge q and momentum k would be written
\begin{equation}
    \ket{\mathrm{in}, \ \mathrm{dressed}} = \lim_{t\rightarrow-\oo}e^{iq\Phi(t,k)}a^{\dagger}_{t}(k)\ket{0} \ .
\end{equation}
Arbitrary dressed states are then written as follows (we again suppress polarization indices):
\begin{equation}
    \begin{aligned}
        &\bra{\mathrm{out}, \ \mathrm{dressed}}=\lim_{t\rightarrow+\oo}\bra{0}a_{t}(p_1)e^{-iq_{1}\Phi(t,p_1)}\cdots b_{t}(p_{n'_s})e^{-iq_{n'_s}\Phi(t,p_{n'_s})}\a_{t}(p_{n'_s+1})\cdots\a_{t}(p_{n'_s+n'_{\gamma}}) \ ,\\
        &\ket{\mathrm{in}, \ \mathrm{dressed}}=\lim_{t\rightarrow-\oo}e^{iq_{1}\Phi(t,k_{1})}a^{\dagger}_{t}(k_1)\cdots e^{iq_{n_s}\Phi(t,k_{n_s})}b^{\dagger}_{t}(k_{n_s})\a^{\dagger}_{t}(k_{n_s+1})\cdots\a^{\dagger}_{t}(k_{n_s+n_{\gamma}})\ket{0} \ .
    \end{aligned}
\end{equation}
The function $\Phi$ is defined as \cite{Kulish:1970ut} \footnote{Here we drop the part known as Dollard's operator which cancels the Coulomb divergence, as it will not be required for our considerations. }
\begin{equation}
\label{FKphidefqed}
    \Phi(t,p)=-i\int\frac{d^3\bk}{(2\pi)^3}\frac{1}{2\w_k}[f(k,p,t)\cdot \e^{r}\a^{\dagger}_{t,r}(k)-f^*(k,p,t)\cdot\e^{*r}(k)\a_{t,r}(k)] \ ,
\end{equation}
where
\begin{equation}
    f_{\mu}(k,p,t) = (\frac{p_{\mu}}{k\cdot p}-\frac{c_{\mu}(k)}{\w_k})\varphi(k,p)e^{it\frac{ p \cdot k}{\w_p}} \ .
\end{equation}
Let us investigate the various parts of this dressing. Rather than deriving everything we simply state the results found in \cite{Kulish:1970ut}. The $p_{\mu}/(k\cdot p)$ factor is exactly what cancels IR divergences. In fact, in order to cure virtual soft divergences in massive scalar scattering (no hard external photons) one need simply use $f_{\mu}(k,p)=p_{\mu}/(k\cdot p)$. The exponential factor $\exp(it (p \cdot k)/\w_p)$ appears whilst deriving the dressing, but it is usually dropped in the literature. This is fine when showing that IR divergences cancel, but in general one might wish to keep the time dependence. \footnote{A time dependent formalism for QED dressings was also constructed in a series of papers by Hirai and Sugishita \cite{Hirai:2019gio, Hirai:2020kzx, Hirai:2022yqw}. It would be interesting to see to what extent our final results agree.} For a discussion on the order of limits in memory calculations and time-dependence see Appendix \ref{AppendixMemCalc}. The function $\varphi$ is any smoothing function and is needed to make certain integrals converge. It satisfies $\varphi=1$ in a neighbourhood of $k=0$ and decays for large $|k|$. Lastly, the factor $c_{\mu}/\w_k$ is required for gauge fixing and restricting to the physical Fock space, i.e. restricting the asymptotic Hilbert space to vectors satisfying the Gupta-Bleuler condition
\begin{equation}
    k_{\mu}\a^{\mu}\Psi=0 \ .
\end{equation}
In particular, $c_{\mu}(k)$ is a function of the null vector $k$ and satisfies
\begin{equation}
    c_{\mu}(k)c^{\mu}(k)=0, \quad k_{\mu}c^{\mu}(k)=\w_k, \quad \forall  \ k \ .
\end{equation}
Equivalently, one can write
\begin{equation}
    c_{\mu}(k)=\frac{\w_k q_{\mu}(k)}{q\cdot k} \ ,
\end{equation}
where $q:S^2\rightarrow S^2$ is a null vector which is a function of the null vector $k$. Then the conditions on $c_{\mu}$ are satisfied if $q_{\mu}$ has no fixed point. In \cite{Kulish:1970ut} it is shown that any choice satisfying these conditions will produce a Fock space with the finite, correct S-matrix elements.\footnote{Again, this statement is true up to issues with the FK dressings as discussed in Subsection \ref{introFK}.} However, we wish to show that in order to correctly capture the memory effect one must choose the (unique rotationally equivariant $q_{\mu}$, up to rescalings which drop out of $c_{\mu}$)
\begin{equation}
    q^{\mu}(k) = (1,-\hat{\bk}) \ .
\end{equation}
Note that whilst rotational equivariance motivates this choice, it is not assumed in the proof of uniqueness. This gives 
\begin{equation}
    c^{\mu}(k)=-\frac{1}{2}(1,-\hat{\bk}) \ .
\end{equation}
To summarise, we claim that the dressing must be chosen to be
\begin{equation}
    \Phi(t,p)=-i\int\frac{d^3\bk}{(2\pi)^3}\frac{1}{2\w_k}[ f(k,p,t)\cdot \e^{r}\a^{\dagger}_{t,r}(k)- f^*(k,p,t)\cdot\e^{*r}(k)\a_{t,r}(k)] \ ,
\end{equation}
where
\begin{equation}
     f_{\mu}(k,p,t) = (\frac{p_{\mu}}{k\cdot p}-\frac{c_{\mu}(k)}{\w_k})\varphi(k,p)e^{it\frac{ p \cdot k}{\w_p}} \ , \quad c^{\mu}(k)=-\frac{1}{2}(1,-\hat{\bk}) \ .
\end{equation}
In particular, we claim that if and only if the dressing is chosen in this manner will it satisfy the requirements laid out in \cite{Kulish:1970ut} and also reproduce the correct classical electromagnetic memory effect due to the outgoing state. To prove this we must first write down the memory operator. We can define the classical electromagnetic memory as \cite{Bieri:2013hqa} 
\begin{equation}
    \Delta^{\mathrm{EM}}_{X}=\int^{\oo}_{-\oo}duE^{(1)}_{X} \ , \ \quad X\in\{z,\bar{z}\} \ ,
\end{equation}
where the $E^{(1)}_{X}(u)$ are the angular components of the electric field due to the source at order $1/r$ in a large $r$ expansion at constant retarded time $u$. One can show that in the absence of charged massless fields \cite{Bieri:2013hqa}
\begin{equation}
\label{classicalqedmemoryresult}
    D^{X}\Delta^{\mathrm{EM}}_{X} = \bigg(E^{(2)}_{r}(\oo)-[E^{(2)}_{r}(\oo)]_{[0]}\bigg)-\bigg(E^{(2)}_r(-\oo)-[E^{(2)}_r(-\infty)]_{[0]}\bigg) \ ,
\end{equation}
where $E^{(2)}_r(u)$ is the radial component at order $1/r^2$ in a large $r$ expansion of the electric field due to the source at constant retarded time $u$. Here we have dropped the null current term since there are no massless charged particles in our theory. We have also defined $[X]_{[0]}$ as the $\ell=0$ part of $X$.

Let's start by showing that our choice of dressing at least reproduces this. To perform the same calculation using scattering theory we can write $\Delta^{\mathrm{EM}}_{X}$ as a detector  \cite{He:2014cra, Gabai:2016kuf, Campiglia:2015qka, Oertel:2026wsm}. For the mathematical details of the definition see Appendix \ref{AppendixDetDef}.  We have
\begin{equation}
\label{qedmemdetdef}
    D^X\Delta^{\mathrm{EM}}_{X}(z,\bar{z}) = -\frac{4\pi}{\gamma_{z\bar{z}}}\lim_{\w_0\rightarrow0}(\partial_z L_{\omega_0}[F_{u\Bar{z}} ](\infty,y)+\partial_{\bar{z}} L_{\omega_0}[F_{uz}](\infty,y) ) \ .
\end{equation}
Here the  stereographic coordinates $\{z,\bar{z}\}$ refer to the direction in which the detector approaches the celestial sphere, and this direction is parametrized by the null vector $y$ where
\begin{equation}
    y=(1, \frac{z+\bar{z}}{1+|z|^2},i\frac{\bar{z}-z}{1+|z|^2},\frac{1-|z|^2}{1+|z|^2}) \ .
\end{equation}
We have also used 
\begin{equation}
\gamma_{z\bar{z}}=\frac{2}{(1+|z|^2)^2} \ .
\end{equation}
To calculate the contribution of a general out state to the memory effect we can act this on a general FK dressed out scattering state defined by
\begin{equation}
    \begin{aligned}
        &\bra{\mathrm{out}, \ \mathrm{dressed}}=\lim_{t\rightarrow+\oo}\bra{0}a_{t}(p_1)e^{-iq_{1}\Phi(t,p_1)}\cdots b_{t}(p_{n'_s})e^{-iq_{n_{s'}}\Phi (t,p_{n'_s})}\a_{t}(p_{n'_s+1})\cdots\a_{t}(p_{n'_s+n'_{\gamma}}) \ .\\
    \end{aligned}
\end{equation}
Note that we expect the memory contribution from the outgoing state to only be the first two terms in (\ref{classicalqedmemoryresult}). A measurement of $E_{r}^{(2)}(-\oo)$ would be performed at a spacetime point which is space-like separated from the scattering event, and thus can only depend on the incoming state. We find \cite{Gabai:2016kuf, Oertel:2026wsm} (see Appendix \ref{AppendixMemCalc} for calculation)
\begin{equation}
    \bra{\mathrm{out}, \ \mathrm{dressed}}D^{X}\Delta^{\mathrm{EM}}_{X}(z,\bar{z})=\bigg(-\sum_{s'=1}^{n'_{s}}q_{s'}-\sum_{s'=1}^{n'_{s}}\frac{q_{s'}p^2_{s'}}{(p_{s'}\cdot y)^2}\bigg)\bra{\mathrm{out}, \ \mathrm{dressed}} \ .
\end{equation}
Indeed this is equal to the classical result in (\ref{classicalqedmemoryresult}) as we show in Appendix \ref{AppendixClassCalc}. In particular, the radial electric field sensed at  $(u,r,z,\bar{z})$ to leading order in $1/r$ near future null infinity due to a massive charged particle with charge $q$ and momentum $p$ is exactly 
\begin{equation}
    E^{(2)}_{r}(\infty)= -\frac{qp^2}{(p\cdot \hat{y})^2}, \quad \hat{y}=(1, \frac{z+\bar{z}}{1+|z|^2},i\frac{\bar{z}-z}{1+|z|^2},\frac{1-|z|^2}{1+|z|^2}) \ .
\end{equation}
This immediately gives
\begin{equation}
    [E^{(2)}_{r}(\oo)]_{[0]}=q_{\mathrm{tot}} \ .
\end{equation}
Thus, 
\begin{equation}
    \bra{\mathrm{out}, \ \mathrm{dressed}}D^{X}\Delta^{\mathrm{EM}}_{X}(z,\bar{z})=\bigg(E^{(2)}_r(\oo)-[E^{(2)}_{r}(\oo)]_{[0]}\bigg)\bra{\mathrm{out}, \ \mathrm{dressed}} \ .
\end{equation}
Thus we have managed to reproduce the divergence of the classical memory due to the out scattering state derived in \cite{Bieri:2013hqa}. To move onto uniqueness we must consider the memory vector $\Delta^{\mathrm{EM}}_{X}$ itself, not only its divergence: $D^X$ annihilates the magnetic-parity (curl) part of $\Delta^{\mathrm{EM}}_{X}$, and different choices of $c_{\mu}$ can give the same divergence while differing in this part. The proof is given in Appendix \ref{AppendixMemCalc}.

We should mention that the $[E^{(2)}_{r}(\oo)]_{[0]}$ term cancels with $[E^{(2)}_{r}(-\oo)]_{[0]}$ by charge conservation when one calculates the full memory by including the contribution from the in state. However, the $c_{\mu}$ part of the dressing is needed to obtain the physical Fock space, and thus must be included.  Note also that the  $\w_0\rightarrow0$ limit must be taken before the $t\rightarrow\oo$ limit as shown in Appendix \ref{AppendixMemCalc}; once it is, the eigenvalue is independent of $t$. However we believe that this is more than just a formality, since the $t$ dependence is required to kill the dressing's leading contribution to the energy of the state (by the Riemann-Lebesgue lemma), and keeping the full $t$ dependence of the dressed states and detectors may provide an interesting framework to calculate sub-leading effects in QFT scattering. 

\section{Perturbative quantum gravity}
Now we will tell an analogous story in a theory consisting of a massive scalar field coupled to perturbative quantum gravity. The action is given by
\begin{equation}
    S = \int d^4x\sqrt{-g}\bigg(\frac{1}{16\pi G}R-\frac{1}{2}g^{\mu\nu}\partial_{\mu}\phi\partial_{\nu}\phi-\frac{1}{2}m^2\phi^2\bigg) \ ,
\end{equation}
which we expand in the weak gravity regime using $g_{\mu\nu}=\eta_{\mu\nu}+\sqrt{32\pi G}h_{\mu\nu}$. Expanding the relevant fields in the interaction picture gives (from now on we set $G=1$)
\begin{equation}
    \begin{aligned}
        &h_{\mu\nu}(x) =\int_{p_0>0}\frac{d^4p}{(2\pi)^{3}}\delta(-p^2)\sum_{r=\pm}(\a_r(p)\e^{*r}_{\mu\nu}(p)e^{ipx}+\a_{r}^{\dagger}(p)\e^{r}_{\mu\nu}(p)e^{-ipx}) \ , \\
        &\phi(x)=\int \frac{d^3p}{(2\pi)^3}\frac{1}{2\w_p}(e^{i p x}a(p) +e^{-ipx}a^{\dagger}(p)) \ .
    \end{aligned}
\end{equation}
We are using the constant time commutation relations
\begin{equation}
    \begin{aligned}
        &[a(p),a^{\dagger}(q)] = 2\w_p (2\pi)^{3}\delta^{3}(\bp-\bq) \ ,\\
        &[\a_r(p),\a_s^{\dagger}(q)] = 2\delta_{rs}\w_p  (2\pi)^{3}\delta^{3}(\bp-\bq) \ .
    \end{aligned}
\end{equation}
In canonical quantisation the canonical scattering Fock spaces are formed by acting with creation or annihilation operators defined on time slices at infinitely large $|t|$, i.e.
\begin{equation}
    \begin{aligned}
        &\bra{\mathrm{out}}=\lim_{t\rightarrow+\oo}\bra{0}a_{t}(p_1)\cdots a_{t}(p_{n'_s})\a_{t}(p_{n'_s+1})\cdots\a_{t}(p_{n'_s+n'_{g}}) \ ,\\
        &\ket{\mathrm{in}}=\lim_{t\rightarrow-\oo}a^{\dagger}_{t}(k_1)\cdots a^{\dagger}_{t}(k_{n_s})\a^{\dagger}_{t}(k_{n_s+1})\cdots\a^{\dagger}_{t}(k_{n_s+n_{g}})\ket{0} \ .
    \end{aligned}
\end{equation}
Here we have suppressed the polarization index on the graviton creation and annihilation operators. The above are an out state with $n'_s$ massive scalar particles and $n'_{g}$ gravitons and an in state with $n_s$ massive scalar particles and $n_{g}$ gravitons.  For ordinary scattering calculations this limit is often taken implicitly, and the ordinary scattering Feynman rules and LSZ formula are derived by assuming this limit exists. In fact, this is not true for theories with long range forces (like QED and gravity) and this leads to IR divergences like the Coulomb phase divergence. There are also soft divergences: if one forms S-matrix elements from an in and out state and takes the soft limit of an external graviton, then the Weinberg soft theorem tells us that we encounter an IR divergence which scales as the inverse of the soft graviton's energy \cite{Weinberg:1995mt}:
\begin{equation}
    \bigg|\lim_{\w_g\rightarrow 0}\braket{\mathrm{out}}{\mathrm{in}}\bigg|\sim \lim_{\w_g\rightarrow0}\bigg|\frac{1}{\w_{g}}\bigg|\sim \oo \ .
\end{equation}
As we saw, similar IR divergences can be cured in QED by using Faddeev-Kulish dressings. Faddeev-Kulish dressings have also been constructed for gravity \cite{Ware:2013zja}. In this case one must dress both external gravitons and scalars, since gravitons self interact. As an example, a dressed scalar one particle in state can be written
\begin{equation}
    \ket{\mathrm{in}, \ \mathrm{dressed}} = \lim_{t\rightarrow-\oo}e^{i\Phi(t,k)}a^{\dagger}_{t}(k)\ket{0} \ ,
\end{equation}
and arbitrary dressed in and out states are written (recall we are suppressing polarization indices):
\begin{equation}
    \begin{aligned}
        &\bra{\mathrm{out}, \ \mathrm{dressed}}=\lim_{t\rightarrow+\oo}\bra{0}\prod_{g'=1}^{n'_g}\a_{t}(p_{g'})e^{-i\Phi(t,p_{g'})}\prod_{s'=1}^{n'_s} a_{t}(p_{s'})e^{-i\Phi(t,p_{s'})}   \ ,\\
        &\ket{\mathrm{in}, \ \mathrm{dressed}}=\lim_{t\rightarrow-\oo}\prod_{s=1}^{n_s} e^{i\Phi(t,p_{s})}a^{\dagger}_{t}(p_{s}) \prod_{g=1}^{n_g}e^{i\Phi(t,p_{g})}\a^{\dagger}_{t}(p_{g}) \ket{0} \ .
    \end{aligned}
\end{equation}
We define the dressing function $\Phi$ as \cite{Ware:2013zja} \footnote{Here we drop the part known as Dollard's operator which cancels the Coulomb divergence, as it will not be required for our considerations. }
\begin{equation}
    \Phi(t,p)=-i\sqrt{8\pi }\int\frac{d^3\bk}{(2\pi)^3}\frac{1}{2\w_k}[f(k,p,t)\cdot \e^{r}\a^{\dagger}_{t,r}(k)-f^*(k,p,t)\cdot\e^{*r}(k)\a_{t,r}(k)] \ ,
\end{equation}
where
\begin{equation}
    f_{\mu\nu}(k,p,t) = (\frac{p_{\mu}p_{\nu}}{k\cdot p}-c_{\mu\nu}(k,p))\varphi(k,p)e^{it\frac{ p \cdot k}{\w_p}}\ .
\end{equation}
Here the $p_{\mu}p_{\nu}/(k\cdot p)$ factor is what removes IR divergences. The function $c_{\mu\nu}$ must satisfy $c_{\mu\nu}=c_{\nu\mu}$ , $c_{\mu\nu}k^{\mu}=p_{\nu}$ , and $c^{\mu\nu}c_{\mu\nu}-(c^{\mu}{}_{\mu})^2/2=0$ . The general ansatz is
\begin{equation}
    c_{\mu\nu}(k,p) = \frac{q_{\mu}p_{\nu}+q_{\nu}p_{\mu}}{q\cdot k}-\frac{k\cdot p}{(q\cdot k)^2}q_{\mu}q_{\nu} \ , 
\end{equation}
where $q(k)$ is a null vector which depends on the null vector $k$. As in the QED case, whilst the normal constraints given in \cite{Ware:2013zja} on $c_{\mu\nu}$ are satisfied if $q^{\mu}(k):S^2\rightarrow S^2$ is any function without a fixed point, we will show that we must choose (up to an overall factor which drops out of $c_{\mu\nu}$)
\begin{equation}
    q^{\mu}(k)=(1,-\hat{\bk}) \ .
\end{equation}
This happens to be the unique rotationally equivariant choice, though this is not assumed in the proof of uniqueness. Next, the function $\varphi$ is any smoothing function needed to make certain integrals converge and satisfies $\varphi=1$ in a neighbourhood of $k=0$ and decays for large $|k|$. Lastly, we note that although the exponential factor $\exp(it (p \cdot k)/\w_p)$ is often dropped one may wish to keep it so that the dressing correctly describes sub-leading effects in a $1/t$ expansion. We will keep the $t$ dependence, and we show that doing so requires us to think carefully about the order of limits in memory calculations. For more on this, see Appendix \ref{AppendixMemCalc}. Our claim (proven in Appendix \ref{uniquenessproof}) is then that if and only if the dressing is chosen as above will it satisfy the requirements laid out in \cite{Ware:2013zja} and also ensure that the dressed Fock space reproduces the classical gravity memory effect due to the outgoing state for all scattering states. This includes the first order contributions (e.g. derived in \cite{Bieri:2013ada}), and higher order terms which include contributions from external hard gravitons (originally suggested by Thorne \cite{Thorne:1992sdb}). To prove this we must first write down the gravitational memory operator. We can define the gravitational memory as 
\begin{equation}
    \Delta^{\mathrm{GR}}_{AB}=\frac{1}{2}\int^{\oo}_{-\oo}duN_{AB} \ , \ \quad A,B\in\{z,\bar{z}\} \ .
\end{equation}
Here $N_{AB}=\partial_u C_{AB}$ is the Bondi News tensor and $C_{AB}$ is the shear. Now, the leading form of the $uu$ component of the Einstein equations in a large $r$ expansion at constant $u$ is given by
\begin{equation}
\partial_u m_B = \frac{1}{4}(D^2_zN^{zz}+D^2_{\Bar{z}}N^{\Bar{z}\Bar{z}})- T_{uu}, \quad T_{uu} \equiv \frac{1}{4}N_{zz}N^{zz} \ .
\end{equation}
Using this we find 
\begin{equation}
    D^AD^B\Delta^{\mathrm{GR}}_{AB}=2m_B(\oo)-2m_B(-\oo)+2\int^{\oo}_{-\oo}du T_{uu} \ .
\end{equation}
Since we have no massless particles besides gravitons, in this equation the first two terms correspond to first order effects which were described in \cite{Bieri:2013ada}, whilst the $T_{uu}$ is a second order term and is the contribution to the stress-energy from the gravitational field itself. We can calculate the memory in scattering by writing the memory  $\Delta^{\mathrm{GR}}_{AB}$ as a detector \cite{Strominger:2014pwa, Oertel:2026wsm}. For the mathematical details of this definition see Appendix \ref{AppendixDetDef}. We write 
\begin{equation}
\begin{aligned}
\label{gravitymemdetdef}
    D^AD^B\Delta^{\mathrm{GR}}_{AB} =& \lim_{\w_0\rightarrow0}\sqrt{8\pi}\bigg(D^2_zL_{\w_0}[\partial_u h^{zz}](\oo,y)+D^2_{\bar{z}} L_{\w_0}[\partial_u h^{\bar{z}\bar{z}}](\oo,y)\bigg) \ .
\end{aligned}
\end{equation}
This is a detector approaching the celestial sphere along a null ray aligned with $y$, where
\begin{equation}
    y=(1, \frac{z+\bar{z}}{1+|z|^2},i\frac{\bar{z}-z}{1+|z|^2},\frac{1-|z|^2}{1+|z|^2}) \ .
\end{equation}
To calculate the contribution of a general out state to the memory effect we can act this on a general FK dressed outgoing scattering state defined by
\begin{equation}
    \begin{aligned}
        &\bra{\mathrm{out}, \ \mathrm{dressed}}=\lim_{t\rightarrow+\oo}\bra{0}\prod_{g'=1}^{n'_g}\a_{t}(p_{g'})e^{-i\Phi(t,p_{g'})}\prod_{s'=1}^{n'_s} a_{t}(p_{s'})e^{-i\Phi(t,p_{s'})}  \ .\\
    \end{aligned}
\end{equation}
It is important to note that when we do this we are only calculating the contribution of the outgoing state to the memory, which cannot include any terms defined for $u<0$. For example, $m_B(-\infty)$ depends only on the incoming state since it would be measured at a point not in the causal future of the scattering process. A calculation gives \cite{Oertel:2026wsm} (see Appendix \ref{AppendixMemCalc} for details)
\begin{equation}
    \bra{\mathrm{out}, \ \mathrm{dressed}}D^{A}D^B\Delta^{\mathrm{GR}}_{AB}(z,\bar{z})=\lambda(\hat{y})\bra{\mathrm{out}, \ \mathrm{dressed}} \ .
\end{equation}
where 
\begin{equation}
    \lambda(\hat{y})=\sum_{g'=1}^{n'_g}(-8\w_{g'}-6p_{g'}\cdot \hat{y}+\frac{8\pi\w_{g'}}{\gamma_{z\bar{z}}}\delta^2(z_{g'}-z_y))+\sum_{s'=1}^{n'_s}(-8\w_{s'}-6p_{s'}\cdot \hat{y}-\frac{2m^4}{(p_{s'}\cdot \hat{y})^3}) \ .
\end{equation}
As required this is exactly the expected contribution to the memory by the outgoing state given in \cite{Bieri:2013ada}, plus the contribution from hard external gravitons emitted in the outgoing state suggested by Thorne \cite{Thorne:1992sdb}. We show this in detail in Appendix \ref{AppendixMemCalc}. In summary, Bieri and Garfinkle find that the outgoing state's contribution is given to first order by 
\begin{equation}
\label{BGmemresult}
    -\bigg(E^{(3)}_{rr}(\infty)-\sum_{i=0,1}[E^{(3)}_{rr}(\oo)]_{[i]}\bigg)+8\pi\bigg(F-\sum_{i=0,1}[F]_{[i]}\bigg) \ .
\end{equation}
Here $E_{rr}=C_{rtrt}$ is the electric part of the Weyl tensor, and $E^{(3)}_{rr}(u)$ is the leading ($1/r^3$) coefficient in a large $r$ expansion at constant retarded time $u$. $F$ is the energy radiated per unit solid angle due to massless particles \textit{which are not gravitons} emitted in the scattering process. Thus for our theory, this would be zero. However, we will see that there is also a contribution of exactly the same form by gravitons, as suggested by Thorne \cite{Thorne:1992sdb}. This does not appear in Bieri and Garfinkle's treatment since they work to first order in perturbation theory. In our treatment, it arises from higher order Feynman diagrams. For more on this see Section \ref{ininsection}. Lastly, $[X]_{[i]}$ denotes the $\ell=i$ part of $X$. For the scattering process we considered, we find
\begin{equation}
    \begin{aligned}
        &E_{rr}^{(3)}(\oo)= \sum_{s'=1}^{n'_s}\frac{2m^4}{(p_{s'}\cdot\hat{y})^3} \ ,
    \end{aligned}
\end{equation}
which immediately gives
\begin{equation}
    [E^{(3)}_{rr}(\oo)]_{[0]}=-\sum_{s'=1}^{n'_s}2\w_{s'} \ , \quad [E^{(3)}_{rr}(\oo)]_{[1]}=-\sum_{s'=1}^{n'_s}6\bp_{s'}\cdot \hat{\by} \ .
\end{equation}
The energy due to hard gravitons radiated to null infinity per unit solid angle is 
\begin{equation}
   F=\sum_{g'=1}^{n'_g}\frac{\w_{g'}}{\gamma_{z\bar{z}}}\delta^2(z_{g'}-z_y) \ ,
\end{equation}
which again gives
\begin{equation}
    [F]_{[0]}=\sum_{g'=1}^{n'_g}\frac{\w_{g'}}{4\pi} \ , \quad [F]_{[1]}=\sum_{g'=1}^{n'_g}\frac{3}{4\pi}\bp_{g'}\cdot \hat{\by} \ .
\end{equation}
This confirms that the eigenvalue of the memory detector is indeed the memory due to the outgoing state
\begin{equation}
    \lambda(\hat{y})=
    -\bigg(E^{(3)}_{rr}(\infty)-\sum_{i=0,1}[E^{(3)}_{rr}(\oo)]_{[i]}\bigg)+8\pi\bigg(F-\sum_{i=0,1}[F]_{[i]}\bigg) \ .
\end{equation}
As in the QED case, we have managed to reproduce $D^AD^B\Delta_{AB}^{\mathrm{GR}}$, showing that our dressings work, but uniqueness will require considering $\Delta_{AB}^{\mathrm{GR}}$. The full proof of uniqueness is given in Appendix \ref{uniquenessproof}.

We should mention that the $[E^{(3)}_{rr}(\oo)]_{[i]}$ and $[F]_{[i]}$ terms cancel by energy and momentum conservation when one calculates the full memory by including the contribution from the incoming state. However, the $c_{\mu\nu}$ part of the dressing is needed to obtain the physical Fock space, and thus must be included. Furthermore, one really does need to subtract the $\ell=0$ and $\ell=1$ parts to obtain the correct contribution, for example see Thorne's result in \cite{Thorne:1992sdb}, so we believe this is a genuine correction to papers which drop the $c_{\mu\nu}$ term.   Note also that the  $\w_0\rightarrow0$ limit must be taken before the $t\rightarrow\oo$ limit as shown in Appendix \ref{AppendixMemCalc}; once it is, the eigenvalue is independent of $t$. However we believe that this is more than just a formality, since the $t$ dependence is required to kill the dressing's leading contribution to the energy of the state (by the Riemann-Lebesgue lemma), and keeping the full $t$ dependence of the dressed states and detectors may provide an interesting framework to calculate sub-leading effects in QFT scattering. 

\section{In-in inclusive memory calculations and Christodoulou non-linear memory}
\label{ininsection}
Consider the full non-linear Christodoulou gravitational memory measured at some point on the celestial sphere due to a scattering event. Gravitational waves emitted during the scattering only contribute to the memory starting at second order in perturbation theory. This is why Bieri and Garfinkle note that their treatment in \cite{Bieri:2013ada} does not include these contributions. In particular, Bieri and Garfinkle find a contribution 
\begin{equation}
    8\pi\bigg(F-\sum_{i=0,1}[F]_{[i]}\bigg) \ ,
\end{equation}
where $F$ is the energy emitted to null infinity due to massless particles which are not gravitons. However, Thorne noted that the contribution from gravitational waves emitted in the scattering process can be written as proportional to the energy per unit solid angle emitted to null infinity by external hard gravitons \cite{Thorne:1992sdb}. In \cite{Oertel:2026wsm} it was found that the memory detector reproduces Bieri and Garfinkle's term but also includes energy from external hard massless gravitons emitted. However, this contribution is still effectively higher order in perturbation theory since attaching an outgoing graviton to a Feynman diagram raises the perturbative order at which that diagram contributes in a weak metric expansion. Noticing this leads us to deduce that the memory detector we have used in this paper is the full memory detector with higher order effects appearing from higher order Feynman diagrams, rather than from additional contributions to the memory detector. Indeed, no more soft emissions are relevant, since S-matrix elements for soft emissions vanish when FK dressed states are used.

Let us make this more clear by considering a specific calculation. Consider a process where two scalars comprise the incoming state $\ket{\mathrm{in}}$. The memory can be computed by doing an inclusive Schwinger-Keldysh in-in calculation and we can write \footnote{Memory correlators like these have previously been investigated in \cite{Moult:2025njc}.}
\begin{equation}
    \langle D^AD^B\Delta_{AB}^{\mathrm{GR}}(z,\bar{z})\rangle = \bra{\mathrm{in}} D^AD^B\Delta_{AB}^{\mathrm{GR}}(z,\bar{z})\ket{\mathrm{in}} \ .
\end{equation}
Here the states $\bra{\mathrm{in}}$ and $\ket{\mathrm{in}}$ live on a constant time slice at large negative $t$ and $D^AD^B\Delta_{AB}^{\mathrm{GR}}(z,\bar{z})$ lives on a constant time slice at large positive $t$  approaching a point on the celestial sphere with angular coordinates $(z,\bar{z})$. We can insert a sum over all possible outgoing states $X$ to calculate this explicitly using scattering amplitudes. We find \footnote{Note that the FK dressed outgoing Fock space elements are eigenvectors of the memory detector. To include the incoming memory in the calculation we move the incoming dressing onto the outgoing state, as in \cite{Choi:2017bna}, so that the states $\ket{X,\mathrm{dressed}}$ carry the combined dressing and the eigenvalue $\lambda_X$ below includes the contribution of the incoming state; the overlaps $\braket{\mathrm{in}}{X,\mathrm{dressed}}$ are then IR finite. We also assume that these states furnish a resolution of the identity on the physical Fock space. Due to the possible issues in Subsection \ref{introFK}, this step is somewhat formal.} 
\begin{equation}
    \langle D^AD^B\Delta_{AB}^{\mathrm{GR}}(z,\bar{z})\rangle = \sum_{X} \int d\Pi_X \braket{\mathrm{in}}{X , \mathrm{dressed}} \bra{X , \mathrm{dressed}}D^AD^B\Delta_{AB}^{\mathrm{GR}}(z,\bar{z})\ket{\mathrm{in}} \ .
\end{equation}
Here $\int d\Pi_X$ is simply integrating over the outgoing momenta subject to on-shell and momentum conservation constraints. The contribution to the memory expectation value from some outgoing state $X$ is then proportional to the S-matrix element for the transition from the incoming state to $X$, integrated against the memory eigenvalue 
\begin{equation}
    \langle D^AD^B\Delta_{AB}^{\mathrm{GR}}(z,\bar{z})\rangle_X \propto \int d\Pi_X \lambda_{X}(z,\bar{z})|\mathcal{M}_{2s\rightarrow X}|^2 \ .
\end{equation}
Here $\langle D^AD^B\Delta_{AB}^{\mathrm{GR}}(z,\bar{z})\rangle_X$ is the $X$ contribution to $\langle D^AD^B\Delta_{AB}^{\mathrm{GR}}(z,\bar{z})\rangle$, and $\lambda_{X}(z,\bar{z})$ is the memory eigenvalue associated with $X$
\begin{equation}
    \bra{X, \ \mathrm{dressed}}D^AD^B\Delta_{AB}^{\mathrm{GR}}(z,\bar{z})=\lambda_{X}(z,\bar{z})\bra{X, \ \mathrm{dressed}} \ .
\end{equation}
Denoting an outgoing state given by $X$ plus one additional outgoing graviton as $X+g'$, we find a higher order contribution to $\langle D^AD^B\Delta_{AB}^{\mathrm{GR}}(z,\bar{z})\rangle$ proportional to
\begin{equation}
    \int d\Pi_{X+g'} \lambda_{X+g'}(z,\bar{z})|\mathcal{M}_{2s\rightarrow X+g'}|^2 \ .
\end{equation}
This is higher order in perturbation theory than the leading order contribution since $|\mathcal{M}_{2s\rightarrow X+g'}|^2$ is higher order than $|\mathcal{M}_{2s\rightarrow X}|^2$. For the classical result, we expect all higher order contributions are organised through the KMOC formalism, though we do not compute them explicitly. \footnote{We can also compute quantum corrections to memory using this method.}

Physically, this occurs since once the physical Fock space is constructed it is IR finite under soft emissions, and soft emissions do not give finite contributions (amplitudes for a soft emission are zero). 

\section{Conclusion}
IR divergences due to soft virtual and emitted photons and gravitons can be resolved by constructing gauge fixed, time dependent Faddeev-Kulish dressed Fock spaces in both QED and perturbative quantum gravity \cite{Kulish:1970ut, Ware:2013zja}. The IR divergences that these dressings remove are controlled by the soft theorem \cite{Weinberg:1995mt}, and in recent years have been connected to asymptotic symmetries and the memory effect \cite{He:2014cra, He:2014laa, Strominger:2013jfa, Strominger:2014pwa, Pasterski:2015zua, Gabai:2016kuf, Campiglia:2015qka, Campiglia:2015kxa, Prabhu:2022zcr, Choi:2017bna}. Building on previous work and extending to external hard gravitons and photons it was shown in \cite{Oertel:2026wsm} that using the unique rotationally equivariant version of these dressings correctly reproduces the classical outgoing memory effect. Here we give a detailed formal discussion and prove that this choice is in fact unique (and that equivariance or smoothness need not be assumed). In particular, we develop a fully time dependent formalism for both QED and gravity which requires putting the Fock spaces on constant time surfaces and handling the asymptotic limits carefully. We believe that the developed formalism is not just a mathematically convenient description, but a step towards fully describing the sub-leading properties of Fock spaces. Doing so resolves enough mathematical inconsistencies to begin to calculate in-in memory correlators, though it should be noted that the FK dressing methodology still contains various issues.

We find various upsides to being more mathematically precise. For example, we find that by allowing us to include external hard gravitons, our formalism may be used to calculate higher order perturbative corrections to the memory effect, also known as the non-linear Christodoulou null memory. We show that these arise due to higher order Feynman diagrams in inclusive in-in calculations. 

There are multiple promising future directions. This formalism originally arose by thinking about asymptotically conserved symmetries in gauge theories in 4 dimensions. It thus would be interesting to extend these results to the sub-leading asymptotically conserved charges in gravity described in the literature \cite{Cachazo:2014fwa, Strominger:2015bla, Lysov:2014csa, Pasterski:2015tva, Hamada:2018vrw, Freidel:2021dfs, Freidel:2021ytz,   Geiller:2024bgf, Cresto:2024fhd, Cresto:2024mne}. It would also be interesting to use the formalism developed here and in \cite{Moult:2025njc} to perform quantum memory detector calculations as was done for gravitational energy detectors in \cite{Herrmann:2024yai}.

\appendix

\section{Definitions of memory detectors}
\label{AppendixDetDef}
Here we define the memory detectors following the notation in \cite{Oertel:2026wsm}.
\subsection{QED memory detector definition}
In (\ref{qedmemdetdef}) we wrote the QED memory detector as 
\begin{equation}
    D^X\Delta^{\mathrm{EM}}_{X}(z,\bar{z}) = -\frac{4\pi}{\gamma_{z\bar{z}}}\lim_{\w_0\rightarrow0}(\partial_z L_{\omega_0}[F_{u\Bar{z}} ](\infty,y)+\partial_{\bar{z}} L_{\omega_0}[F_{uz}](\infty,y) ) \ ,
\end{equation}
where
\begin{equation}
    y=(1, \frac{z+\bar{z}}{1+|z|^2},i\frac{\bar{z}-z}{1+|z|^2},\frac{1-|z|^2}{1+|z|^2}) \ .
\end{equation}
Let us unpack this. First, given the field $F_{u\bar{z}}$, we can send it to future null infinity in the null direction $y$ using
\begin{equation}
    F_{u\bar{z}}(u,y)=\lim_{L\rightarrow\oo} F_{u\bar{z}}(x+Ly) \ ,
\end{equation}
where $u=-y\cdot x$. This can be evaluated in practice by expanding into momentum modes and performing a stationary phase approximation. Next, we define a $\omega_{0}$-deformed light transform by defining
\begin{equation}
    L_{\omega_0}[F_{u\Bar{z}} ](\infty,y) =\frac{1}{2}\int du (e^{i\w_0 u}+e^{-i\w_0 u})F_{u\bar{z}}(u,y) \ .
\end{equation}
In order to probe the zero mode (which is the memory), we will take $\w_0\rightarrow 0$. We discuss the limits involved in more detail in the calculation in Appendix \ref{AppendixMemCalc}. Here we will just note that the limits are to be taken in the following order: $L\rightarrow\oo$ first, and then $\w_0\rightarrow 0$, and finally $t\rightarrow \oo$. Both the detector and the outgoing state are then defined on a fixed time slice at time $t$.

\subsection{Perturbative quantum gravity memory detector definition}
In (\ref{gravitymemdetdef}) we wrote the gravitational memory detector as
\begin{equation}
\begin{aligned}
    D^AD^B\Delta^{\mathrm{GR}}_{AB} =& \lim_{\w_0\rightarrow0}\sqrt{8\pi}\bigg(D^2_zL_{\w_0}[\partial_u h^{zz}](\oo,y)+D^2_{\bar{z}} L_{\w_0}[\partial_u h^{\bar{z}\bar{z}}](\oo,y)\bigg) \ .
\end{aligned}
\end{equation}
where
\begin{equation}
    y=(1, \frac{z+\bar{z}}{1+|z|^2},i\frac{\bar{z}-z}{1+|z|^2},\frac{1-|z|^2}{1+|z|^2}) \ .
\end{equation}
Here we have defined
\begin{equation}
    L_{\omega_0}[\partial_{u}{h^{zz}} ](\infty,y) =\frac{1}{2}\int du (e^{i\w_0 u}+e^{-i\w_0 u})\partial_{u}h^{zz}(u,y) \ .
\end{equation}
where 
\begin{equation}
    h^{zz}(u,y)=\lim_{L\rightarrow\oo}L^{3}h^{zz}(x+Ly) \ .
\end{equation}
This limit sends $h^{zz}(u,y)$ to the celestial sphere in the null direction $y$. The factor of $L^{3}$ is to ensure finiteness in the limit. It is important to consider this detector as a distribution valued operator. Technically, all limits are to be taken after integration against a test function of external momentum. For our purposes the $L\rightarrow\oo$ limit can be taken first using the stationary phase approximation. Either way, the order of limits is as follows: $L\rightarrow\oo$ should be taken first, and then $\w_0\rightarrow0$, and finally $t\rightarrow\oo$. The detector and outgoing state are then effectively defined on a fixed time slice at time $t$. This is made clearer in Appendix \ref{AppendixMemCalc} where the explicit calculation of the memory is performed.

\section{Detector memory calculations}
\label{AppendixMemCalc}
Here we will partially go through the calculations of the eigenvalues of the memory detectors acting on dressed states. In particular, we focus on the reasoning which shows that the dressings must be chosen in a unique way. 
\subsection{QED detector memory calculation}
We wish to show
\begin{equation}
    \bra{\mathrm{out}, \ \mathrm{dressed}}D^{X}\Delta^{\mathrm{EM}}_{X}(z,\bar{z})=\bigg(-\sum_{s'=1}^{n'_{s}}q_{s'}-\sum_{s'=1}^{n'_{s}}\frac{q_{s'}p^2_{s'}}{(p_{s'}\cdot \hat{y})^2}\bigg)\bra{\mathrm{out}, \ \mathrm{dressed}} \ ,
\end{equation}
where
\begin{equation}
    \begin{aligned}
        &\bra{\mathrm{out}, \ \mathrm{dressed}}=\lim_{t\rightarrow+\oo}\bra{0}a_{t}(p_1)e^{-iq_{1}\Phi(t,p_1)}\cdots b_{t}(p_{n'_s})e^{-iq_{n'_s}\Phi (t,p_{n'_s})}\a_{t}(p_{n'_s+1})\cdots\a_{t}(p_{n'_s+n'_{\gamma}}) \ .\\
    \end{aligned}
\end{equation}
We will summarise this calculation as it appears in \cite{Oertel:2026wsm}. This calculation comes down to computing the commutator of the memory detector with the FK dressing, and the photon creation and annihilation operators, and then substituting. We take the memory operator to annihilate the vacuum, since the stationary phase approximation shows that it creates null photons, and dressed states have no IR divergences to cancel the associated zeros. Recall that the dressing function depends on a gauge fixing function $c_{\mu}$ and is given by 
\begin{equation}
    \Phi(t,p)=-i\int\frac{d^3\bk}{(2\pi)^3}\frac{1}{2\w_k}[f(k,p,t)\cdot \e^{r}\a^{\dagger}_{t,r}(k)-f^*(k,p,t)\cdot\e^{*r}(k)\a_{t,r}(k)] \ , 
\end{equation}
where 
\begin{equation}
    f_{\mu}(k,p,t) = (\frac{p_{\mu}}{k\cdot p}-\frac{c_{\mu}(k)}{\w_k})\varphi(k,p)e^{it\frac{ p \cdot k}{\w_p}} \ ,
\end{equation}
and where $c_{\mu}(k)$ satisfies
\begin{equation}
    c_{\mu}(k)c^{\mu}(k)=0, \quad k_{\mu}c^{\mu}(k)=\w_k, \quad \forall  \ k \ .
\end{equation}
Computing the relevant commutators gives \cite{Oertel:2026wsm}
\begin{equation}
    \begin{aligned}
        &[D^{X}\Delta^{\mathrm{EM}}_{X}(z,\bar{z}),\alpha_{t,r}(k)]=0 \ ,\\
        &[D^{X}\Delta^{\mathrm{EM}}_{X}(z,\bar{z}),a(k)]=0 \ ,\\
        &[D^{X}\Delta^{\mathrm{EM}}_{X}(z,\bar{z}),b(k)]=0 \ ,\\
        &[D^{X}\Delta^{\mathrm{EM}}_{X}(z,\bar{z}),iq_{s'}\Phi(t,p_{s'})]=-q_{s'}-\frac{q_{s'}p^2_{s'}}{(p_{s'}\cdot \hat{y})^2} \ . 
    \end{aligned}
\end{equation}
This reproduces the required result for the choice $c^{\mu}(k)=-\frac{1}{2}(1,-\hat{\bk})$. For a general $c_{\mu}$ satisfying the constraints, before simplification one obtains
\begin{equation}
\label{generalqedmem}
    [D^{X}\Delta^{\mathrm{EM}}_{X}(z,\bar{z}),iq_{s'}\Phi(t,p_{s'})]=-\lambda q_{s'}-\frac{q_{s'}p^2_{s'}}{(p_{s'}\cdot \hat{y})^2} \ ,
\end{equation}
where 
\begin{equation}
    \lambda = \frac{1}{\gamma_{z\bar{z}}}\bigg(2\partial_{z}\bigg[c(z,\bar{z})\cosh({it\w_0\frac{p_{s'}\cdot \hat{y}}{\w_p}})\bigg]+2\partial_{\bar{z}}\bigg[c^*(z,\bar{z})\cosh({it\w_0\frac{p_{s'}\cdot \hat{y}}{\w_p}})\bigg]\bigg) \ ,
\end{equation}
where
\begin{equation}
    c(z,\bar{z}) = \frac{\sqrt{\gamma_{z\bar{z}}}c_{\mu}(\hat{y})\cdot \e^2(z,\bar{z})}{2} \ ,
\end{equation}
and 
\begin{equation}
    \e^2=\frac{1}{\sqrt{2}}(z,1,i,-z)  \ , \quad y=(1, \frac{z+\bar{z}}{1+|z|^2},i\frac{\bar{z}-z}{1+|z|^2},\frac{1-|z|^2}{1+|z|^2}) \ . 
\end{equation}
Firstly, recall the three limits involved which were $L \rightarrow \oo$, $\w_0\rightarrow 0$ and $t\rightarrow\oo$. We took the $L\rightarrow \oo$ first to obtain the above results. The FK dressings encode the memory, and are defined at finite time, and the limit $\w_0\rightarrow 0$ is selecting the zero mode to isolate the desired quantity. We thus must simply take $\w_0\rightarrow 0$ before $t\rightarrow+\oo$. This ordering of the three limits reflects the classical situation, where $r\rightarrow \oo$ is taken before $u\rightarrow \oo$, and the result is the leading order correction to the field which becomes exact as $t\rightarrow \oo$!

We stress that this ordering choice is not an ad hoc choice and is justified physically. It is also the choice which gives conservation of large gauge charges to leading order if one takes into account the time dependence. Note that this is ultimately equivalent to simply dropping the time-dependence for leading order memory calculations.

Next, the unique rotationally equivariant choice of $c_{\mu}$ satisfying the constraints is
\begin{equation}
    c^{\mu}(k)=-\frac{1}{2}(1,-\hat{\bk}) \ .
\end{equation}
This gives
\begin{equation}
    \partial_z c(z,\bar{z})=\frac{\gamma_{z\bar{z}}}{4} \ \implies  \ \lambda =1 \  ,
\end{equation}
as required.

\subsection{Proof of uniqueness of gauge fixing function $c_{\mu}$ for QED FK dressings }
Now, $\lambda=1$ alone does not single out the above choice. Consider a general choice 
\begin{equation}
q^{\mu}=(1,-\hat{\bv}) \ ,
\end{equation}
 where $\hat{\bv}(\hat{\bk})$ is a unit vector depending on $\bk$ (and never satisfies $\bv=-\bk$ as required in \cite{Kulish:1970ut}), and define
\begin{equation}
 W=\bv_{\perp}/(1+\hat{\bk}\cdot\hat{\bv}) \ ,
\end{equation}
where $\bv_{\perp}$ is the part of $\hat{\bv}$ perpendicular to $\hat{\bk}$. Importantly, this implies that $q^{\mu}\propto(1,-\hat{\bk})$ if and only if $W=0$. Now, redoing the calculation above gives $\lambda=1+\nabla_A W^A$, so any divergence-free $W$ also gives $\lambda=1$: the divergence $D^X$ annihilates the magnetic-parity part of the memory vector. However, the memory vector itself does fix $c_{\mu}$. Its component $\Delta^{\mathrm{EM}}_{\bar{z}}=-4\pi\lim_{\w_0\rightarrow0}L_{\w_0}[F_{u\bar{z}}](\oo,y)$ has, on a dressed out state, the eigenvalue
\begin{equation}
    \Delta^{\mathrm{EM}}_{\bar{z}}=\sum_{s'=1}^{n'_s}q_{s'}\Big(\partial_{\bar{z}}\log(-p_{s'}\cdot\hat{y})-W_{\bar{z}}\Big) \ , \qquad W_{\bar{z}}\equiv W\cdot\partial_{\bar{z}}\hat{\by} \ ,
\end{equation}
and similarly for $\Delta^{\mathrm{EM}}_{z}$. The $W$-independent part is exactly the classical memory of the outgoing charges, which is a pure gradient (it has no magnetic part). Thus, the other part immediately implies that $W$ must vanish everywhere. Indeed, the only dependence of the expression above on $\hat{\bv}(\hat{\bk})$ in the direction $\hat{y}$ is at $\hat{\bk}=\hat{\by}$, so obtaining the correct memory for any outgoing state with non-zero charge requires $W=0$ at every point of the celestial sphere, i.e. $c^{\mu}(k)=-\frac{1}{2}(1,-\hat{\bk})$. The analogous statement in gravity is proven in Appendix \ref{uniquenessproof}.
\subsection{Perturbative quantum gravity detector memory calculation}
We wish to show 
\begin{equation}
    \bra{\mathrm{out}, \ \mathrm{dressed}}D^{A}D^B\Delta^{\mathrm{GR}}_{AB}(z,\bar{z})=\lambda(\hat{y})\bra{\mathrm{out}, \ \mathrm{dressed}} \ .
\end{equation}
where 
\begin{equation}
    \lambda(\hat{y})=\sum_{g'=1}^{n'_g}(-8\w_{g'}-6p_{g'}\cdot \hat{y}+\frac{8\pi\w_{g'}}{\gamma_{z\bar{z}}}\delta^2(z_{g'}-z_y))+\sum_{s'=1}^{n'_s}(-8\w_{s'}-6p_{s'}\cdot \hat{y}-\frac{2m^4}{(p_{s'}\cdot \hat{y})^3}) \ .
\end{equation}
The outgoing state is completely general and given by
\begin{equation}
    \begin{aligned}
        &\bra{\mathrm{out}, \ \mathrm{dressed}}=\lim_{t\rightarrow+\oo}\bra{0}\prod_{g'=1}^{n'_g}\a_{t}(p_{g'})e^{-i\Phi(t,p_{g'})}\prod_{s'=1}^{n'_s} a_{t}(p_{s'})e^{-i\Phi(t,p_{s'})}  \ .\\
    \end{aligned}
\end{equation}
Recall that the dressing function depends on a gauge fixing function $c_{\mu\nu}$ and is given by 
\begin{equation}
    \Phi(t,p)=-i\sqrt{8\pi }\int\frac{d^3\bk}{(2\pi)^3}\frac{1}{2\w_k}[f(k,p,t)\cdot \e^{r}\a^{\dagger}_{t,r}(k)-f^*(k,p,t)\cdot\e^{*r}(k)\a_{t,r}(k)] \ ,
\end{equation}
where
\begin{equation}
    f_{\mu\nu}(k,p,t) = (\frac{p_{\mu}p_{\nu}}{k\cdot p}-c_{\mu\nu}(k,p))\varphi(k,p)e^{it\frac{ p \cdot k}{\w_p}}\ .
\end{equation}
We will summarise this calculation as it appears in \cite{Oertel:2026wsm}. This calculation comes down to computing the commutator of the memory detector with the FK dressing, and the graviton creation and annihilation operators, and then substituting. Recall that we defined
\begin{equation}
\begin{aligned}
    D^AD^B\Delta^{\mathrm{GR}}_{AB} =& \lim_{\w_0\rightarrow0}\sqrt{8\pi}\bigg(D^2_zL_{\w_0}[\partial_u h^{zz}](\oo,y)+D^2_{\bar{z}} L_{\w_0}[\partial_u h^{\bar{z}\bar{z}}](\oo,y)\bigg) \ ,
\end{aligned}
\end{equation}
where the stationary phase approximation gives
\begin{equation}
    \begin{aligned}
        &L_{\w_0}[\partial_{u}h^{zz}](\oo,y) =- \frac{|\w_0|(1+|z|^2)^2}{16\pi}(\a^{\dagger}_{t,1}(|\w_0| y)+\a_{t,2}(|\w_0| y)) \ ,\\
        &L_{\w_0}[\partial_{u}h^{\bar{z}\bar{z}}](\oo,y) =  -\frac{|\w_0|(1+|z|^2)^2}{16\pi}( \a^{\dagger}_{t,2}(|\w_0| y)+\a_{t,1}(|\w_0| y)) \ .\\
    \end{aligned}
\end{equation}
First, note that computing the commutator of $D^AD^B\Delta^{\mathrm{GR}}_{AB}$ with operators defined at time $t$ requires paying attention to the order of the three limits involved which are $L \rightarrow \oo$, $\w_0\rightarrow 0$ and $t\rightarrow\oo$. We will take the $L\rightarrow \oo$ first and use the stationary phase approximation to obtain the below results. The FK dressings encode the memory, and are defined at finite time, and the limit $\w_0\rightarrow 0$ is selecting the zero mode to isolate the desired quantity. We thus must simply take $\w_0\rightarrow 0$ before $t\rightarrow\oo$. This ordering of the three limits reflects the classical situation, where $r\rightarrow \oo$ is taken before $u\rightarrow \oo$, and the result is the leading order correction to the field which becomes exact as $t\rightarrow \oo$!

We stress that this is not an ad hoc choice and is justified physically. It is also the choice which gives conservation of large gauge charges to leading order if one takes into account the time dependence. Note that this is ultimately equivalent to simply dropping the time-dependence, but we are simply trying to understand how memory may be calculated using time dependent Fock spaces.

Computing the relevant commutators then gives 
\begin{equation}
    \begin{aligned}
        &[D^{A}D^B\Delta^{\mathrm{GR}}_{AB}(z,\bar{z}),\alpha_{t,r}(k)]=0 \ ,\\
        &[D^{A}D^B\Delta^{\mathrm{GR}}_{AB}(z,\bar{z}),a_t(k)]=0 \ ,\\
        &[D^{A}D^B\Delta^{\mathrm{GR}}_{AB}(z,\bar{z}),i\Phi(t,p_{s'})]= \Lambda(p_{s'},c)-\frac{2m^4}{(p_{s'}\cdot \hat{y})^3}\ , \\ 
        &[D^{A}D^B\Delta^{\mathrm{GR}}_{AB}(z,\bar{z}),i\Phi(t,p_{g'})]= \Lambda(p_{g'},c)+ \frac{8\pi\w_{g'}}{\gamma_{z\bar{z}}}\delta^2(z_{g'}-z_y) \ . \\ 
    \end{aligned}
\end{equation}
Here we have defined
\begin{equation}
    \Lambda(p,c) = 2\bigg(D^2_{z}\bigg[c(z,\bar{z})\cosh({it\w_0\frac{p\cdot \hat{y}}{\w_p}})\bigg]+D^2_{\bar{z}}\bigg[c^*(z,\bar{z})\cosh({it\w_0\frac{p\cdot \hat{y}}{\w_p}})\bigg]\bigg) \ ,
\end{equation}
where
\begin{equation}
    c(z,\bar{z})=\frac{1}{\gamma_{z\bar{z}}}\bigg(\frac{q_{\mu}p_{\nu}+q_{\nu}p_{\mu}}{q\cdot \hat{y}}-\frac{\hat{y}\cdot p}{(q\cdot \hat{y})^2}q_{\mu}q_{\nu}\bigg)\e^{2,\mu\nu} \ .
\end{equation}
Here the graviton polarization tensor is given by $\e^{2,\mu\nu}=\e^{2,\mu}\e^{2,\nu}$ where
\begin{equation}
    \e^2=\frac{1}{\sqrt{2}}(z,1,i,-z) \ .
\end{equation}
We are also using
\begin{equation}
    y=(1, \frac{z+\bar{z}}{1+|z|^2},i\frac{\bar{z}-z}{1+|z|^2},\frac{1-|z|^2}{1+|z|^2}) \ .
\end{equation}
Taking the limit $\w_0\rightarrow 0$ as we discussed and using the unique rotationally equivariant choice
\begin{equation}
    q^{\mu}(k)=(1,-\hat{\bk}) \ ,
\end{equation}
we find
\begin{equation}
    \Lambda(p,c)=-8\w_p-6p\cdot \hat{y} \ ,
\end{equation}
as desired. To finish off we prove that no other $q(k)$ gives the correct memory.
\subsection{Proof of uniqueness of gauge fixing function $c_{\mu\nu}$ for gravitational FK dressings }
\label{uniquenessproof}
The statement we wish to prove is that given the FK dressed states discussed, the memory eigenvalue will only match the classical memory if the gauge fixing function $c_{\mu\nu}(k,p)$ is chosen in a unique manner, which is given by
\begin{equation}
    c_{\mu\nu}(k,p) = \frac{q_{\mu}p_{\nu}+q_{\nu}p_{\mu}}{q\cdot k}-\frac{k\cdot p}{(q\cdot k)^2}q_{\mu}q_{\nu} \ , \quad  q^{\mu}(k)=(1,-\hat{\bk}) \ .
\end{equation}
In the previous subsection we showed that this choice reproduces the double divergence of the memory. Now we will show that it is the unique correct choice by considering the full memory tensor. First, note that the above ansatz is completely general up to terms which drop out when contracted with physical polarizations. Thus, we can prove uniqueness of $q^{\mu}(k)$ assuming this ansatz and the condition that $q^{\mu}(k)$ is a null vector with no fixed point. To do this, it is easiest to work in a different gauge. So far we have been working in a gauge where the gravitational polarization vectors are given by $\e^{i,\mu\nu}=\e^{i,\mu}\e^{i,\nu}$ where
\begin{equation}
    \e^1=\frac{1}{\sqrt{2}}(\bar{z},1,-i,-\bar{z}) \ , \quad \e^2=\frac{1}{\sqrt{2}}(z,1,i,-z) \ .
\end{equation}
Let us now instead work in the gauge used in \cite{Ware:2013zja}, i.e. choose polarization vectors given by the gauge transformation
\begin{equation}
\e^1\rightarrow \e^1-\frac{1}{\sqrt{2}}\bar{z}\hat{k} \ , \quad \e^2\rightarrow \e^2-\frac{1}{\sqrt{2}}z\hat{k} \ .
\end{equation}
Now the polarization vectors are purely spatial, and it is easy to check that with the choice $ q^{\mu}(k)=(1,-\hat{\bk})$ the new polarization basis satisfies
\begin{equation}
\e^1\cdot q = \e^2\cdot q = 0 \ .
\end{equation}
This implies that we now have 
\begin{equation}
\e^{1,\mu\nu} c_{\mu\nu} = \e^{2,\mu\nu} c_{\mu\nu} = 0 \ .
\end{equation}
Of course, the memory due to the outgoing state should be gauge independent, and indeed it is.  In this choice of gauge the correct memory is still recovered even though there are no contributions from $c_{\mu\nu}(k,p)$, as is easily checked by calculating the memory eigenvalue again in this gauge. The entire contribution is now from the $p_{\mu}p_{\nu}/(k\cdot p)$ term! It is now easy to finish the proof, provided we compare the memory tensor $\Delta^{\mathrm{GR}}_{AB}$ itself rather than only its double divergence. The detector (\ref{gravitymemdetdef}) is built from the components
\begin{equation}
    \Delta^{zz}(z,\bar{z})=\lim_{\w_0\rightarrow0}\sqrt{8\pi}\,L_{\w_0}[\partial_u h^{zz}](\oo,y) \ , \qquad D^AD^B\Delta^{\mathrm{GR}}_{AB}=D^2_z\Delta^{zz}+D^2_{\bar{z}}\Delta^{\bar{z}\bar{z}} \ ,
\end{equation}
and by the stationary phase result above the dressed out states are eigenstates of $\Delta^{zz}$ and $\Delta^{\bar{z}\bar{z}}$ themselves, with eigenvalues that are local on the celestial sphere. As before we can prove uniqueness using a pointwise argument since the memory must be reproduced for all possible states and the eigenvalue in the direction $\hat{y}$ depends on $c_{\mu\nu}(k,p)$ only at $\hat{\bk}=\hat{\by}$. Comparing the full tensor matters, because $D^AD^B$ annihilates the magnetic-parity part of a symmetric traceless tensor on the sphere. 

First, note that the previous choice $q^{\mu}(k)=(1,-\hat{\bk})$ does reproduce the full tensor. We can check this by first noting that the classical result is purely electric (e.g. see \cite{Winicour:2014ska, Madler:2016ggp, Bieri:2018asm}) and that $D^AD^B$ is invertible on electric parity tensors with $\ell\geq 2$. Then it remains to check that the eigenvalue of $\Delta_{\bar{z}\bar{z}}$ is purely electric. Indeed, the eigenvalue is proportional to $(\partial_{\bar{z}}f)^2/f$ with $f=p\cdot\hat{y}$, and since $D^2_{\bar{z}}f=0$ one has $(\partial_{\bar{z}}f)^2/f=D^2_{\bar{z}}(f\log (-f)-f)$, so the eigenvalue tensor is of purely electric parity. Now let $q^{\mu}(k)$ be any null vector without a fixed point. Since $c_{\mu\nu}$ is invariant under $q\rightarrow f(k)q$, we may write 
\begin{equation}
q^{\mu}=(1,-\hat{\bv}) \ ,
\end{equation}
where $\hat{\bv}(\hat{\bk})$ a unit vector depending on $\bk$ (and which never satisfies $\hat{\bv}=-\hat{\bk}$), and define 
\begin{equation}
W=\bv_{\perp}/(1+\hat{\bk}\cdot\hat{\bv}) \ ,
\end{equation}
where $\bv_{\perp}$ is the part of $\hat{\bv}$ perpendicular to $\hat{\bk}$. Importantly, this implies that $q^{\mu}\propto(1,-\hat{\bk})$ if and only if $W=0$. In the gauge above the polarization vector is $\e^{2}=(0,\partial_{\bar{z}}\hat{\by})/\sqrt{\gamma_{z\bar{z}}}$, and a short calculation gives, at $k=\w_k\hat{y}$,
\begin{equation}
    \e^{2,\mu\nu}c_{\mu\nu}(k,p)=\frac{2W_{\bar{z}}\,\partial_{\bar{z}}(p\cdot\hat{y})-(p\cdot\hat{y})\,W_{\bar{z}}^{2}}{\gamma_{z\bar{z}}\,\w_k} \ , \qquad W_{\bar{z}}\equiv W\cdot\partial_{\bar{z}}\hat{\by} \ ,
\end{equation}
together with its complex conjugate for $\e^{1}$. Since $c_{\mu\nu}$ contributes nothing in this gauge for $q^{\mu}=(1,-\hat{\bk})$, the dressing reproduces the classical memory tensor for every outgoing state if and only if this expression vanishes for all $p$ at every point of the celestial sphere. For an outgoing massive particle at rest, $\partial_{\bar{z}}(p\cdot\hat{y})=0$ and $p\cdot\hat{y}=-m$, so we need $m\,W_{\bar{z}}^2=0$, i.e. $W=0$. Thus we must have $q^{\mu}\propto(1,-\hat{\bk})$. Similarly, for an outgoing massless particle, if one requires $2W_{\bar{z}}\,\partial_{\bar{z}}(p\cdot\hat{y})-(p\cdot\hat{y})\,W_{\bar{z}}^{2}=0$, the solutions are $W=0$ and $W_{\bar{z}}=2\partial_{\bar{z}}\log (-p\cdot \hat{y})$. The latter blows up as $p\cdot \hat{y}\rightarrow 0$ (which is impossible if we enforce that $q$ has no fixed points), so the only remaining option is $W=0$. Both arguments are pointwise on the celestial sphere and require neither rotational equivariance nor smoothness of $q$, and so we are done.

\section{Classical memory calculations}
\label{AppendixClassCalc}
Here we derive some classical results that we use
\subsection{EM classical memory calculation}
\begin{figure}[h!]
\centering
\begin{tikzpicture}[
    every node/.style={font=\small},
    wavy/.style={decorate, decoration={snake, amplitude=1.2pt, segment length=5pt}},
]

\foreach \a in {0,45,...,315} { \draw[thick] (0,0) -- ++(\a:0.3); }
\foreach \a in {22.5,67.5,...,337.5} { \draw[thick] (0,0) -- ++(\a:0.18); }

\draw[-{Stealth[length=5pt]}, thick] (-0.7, -1.3) -- (-0.12, -0.25);
\draw[-{Stealth[length=5pt]}, thick] (0.5, -1.3) -- (0.08, -0.25);
\node[circle, draw, inner sep=2pt] at (-0.85, -1.5) {};
\node[circle, draw, inner sep=2pt] at (0.65, -1.5) {};
\node[left=2pt] at (-1.1, -1.5) {in state};


\coordinate (outstate) at (0, 3.1);
\node[circle, draw, inner sep=2pt] at (outstate) {};
\node[left=2pt] at (-0.25, 3.1) {out state};

\draw[-{Stealth[length=5pt]}, thick] (0, 0.35) -- (0, 2.85);
\node[left=3pt] at (0, 1.6) {$\vec{x}=(R+\delta t)\,\vec{v}$};

\draw[-{Stealth[length=5pt]}, thick] (0.2, 3.1) -- (5.8, 0.2);
\node[above right=1pt] at (3.0, 1.65) {$\vec{x}_0-\vec{x}$};




\draw[-{Stealth[length=5pt]}, thick] (0.35, 0) -- (5.8, 0);
\node[above=3pt] at (3.1, 0) {$\vec{x}_0=R\hat{\by}$};

\draw[{Stealth[length=4pt]}-{Stealth[length=4pt]}] (0.35, -0.45) -- (5.8, -0.45);
\node[below=1pt] at (3.1, -0.45) {$R$};

\draw[thick] (5.88, 0.18) -- (5.88, -0.18);
\draw[thick] (5.88, 0.1) -- (6.06, 0.1);
\draw[thick] (5.88, -0.1) -- (6.06, -0.1);
\draw[thick] (6.06, -0.28) rectangle (6.46, 0.28);

\node[right=8pt, align=left] at (6.55, 0) {detector};

\end{tikzpicture}
\caption{A diagram showing a real experiment measuring $E_{r}(u)$ at a large distance $R$ from the origin in the limit $u\rightarrow\oo$, where the scattering event takes place at the origin $r=t=0$. Practically, our order of limits implies that the measurement $E_r(\oo)$ is taken at some time $R/c+\delta t$ where $\delta t$ is large enough to reach a steady state but $ \delta t \ll R/c$. A measurement of $E_r(-\oo)$ is not causally related to the out state and depends only on the in state.}
\label{EMfigure}
\end{figure}
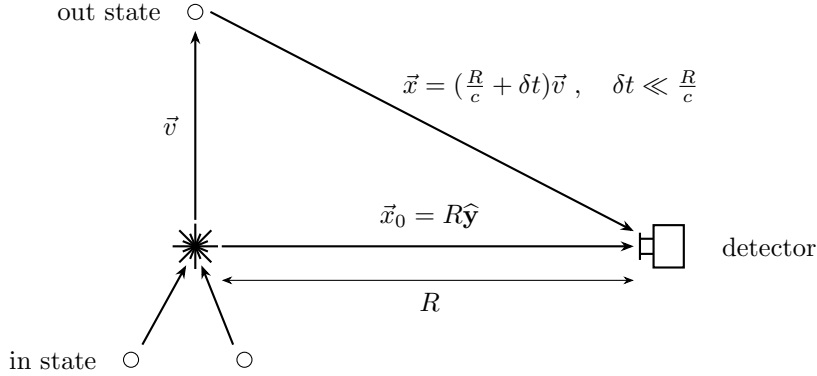

Recall that according to \cite{Bieri:2013hqa} the contribution to the memory from the outgoing state is
\begin{equation}
    D^{X}\Delta^{\mathrm{EM}}_{X} = E^{(2)}_{r}(\oo)- [E^{(2)}_{r}(\oo)]_{[0]} \ .
\end{equation}
We must show that the radial component of the electric field sensed at a point in a direction $y=(1,\hat{\by})$ (to first order in the limit of large $r$ with $u$ constant and approaching $u=+\oo$) due to a massive particle with charge $q$ and momentum $p$ satisfies 
\begin{equation}
    E^{(2)}_{r}(\oo)=-\frac{qp^2}{(p\cdot \hat{y})^2} \ , \qquad E_r=\frac{E^{(2)}_r}{r^2}+\mathcal{O}(r^{-3}) \ .
\end{equation}
Note the order of limits here, we take the limit $r\rightarrow \oo$ at constant $u$ first, and then we take $u\rightarrow\oo$. It is important to consider this in the context of a real experiment. Let the scattering event happen at $r=t=0$, and the detector be distance $R$ from the origin. Then our limits imply that the measurement $E_r(\oo)$ is taken long enough after the scattering that the particles are free, let us say at time $R/c+\delta t$, but not so long that $\delta t$ becomes greater than $R/c$. We illustrate this scenario in Figure \ref{EMfigure}.

Then, using the labelling shown and setting $c=1$, Heaviside's formula for the electric field due to a moving charge with momentum $p$ gives
\begin{equation}
   E^{(2)}_r(\oo)= \lim_{\delta t\rightarrow \oo}\lim_{R\rightarrow\oo}\frac{q R^2(1-\vec{v}^2)\,(\vec{x}_0-\vec{x})\cdot\hat{\by}}{\big((1-\vec{v}^2)|\vec{x}_0-\vec{x}|^2+(\vec{v}\cdot (\vec{x}_0-\vec{x}))^2\big)^{3/2}}=\frac{q(1-\vec{v}^2)}{(1-\vec{v}\cdot\hat{\by})^2}=-\frac{qp^2}{(p\cdot \hat{y})^2} \ , 
\end{equation}
where
\begin{equation}
\vec{x}=(R+\delta t)\,\vec{v} \ , \quad \vec{x}_0=R\,\hat{\by} \ .
\end{equation}
\subsection{GR classical memory calculation}
Here we wish to show that the eigenvalue $\lambda(\hat{y})$ which we obtain by acting the memory detector on an outgoing state is equal to the contribution to the memory due to the outgoing state as derived by Bieri and Garfinkle in \cite{Bieri:2013ada}, plus a higher order contribution from external gravitons. Recall that we found 
\begin{equation}
    \lambda(\hat{y})=\sum_{g'=1}^{n'_g}(-8\w_{g'}-6p_{g'}\cdot \hat{y}+\frac{8\pi\w_{g'}}{\gamma_{z\bar{z}}}\delta^2(z_{g'}-z_y))+\sum_{s'=1}^{n'_s}(-8\w_{s'}-6p_{s'}\cdot \hat{y}-\frac{2m^4}{(p_{s'}\cdot \hat{y})^3}) \ .
\end{equation}
Here the sums over $g'$ and $s'$ correspond to summing over external outgoing hard gravitons and scalars respectively. In Equation (\ref{BGmemresult}) we wrote Bieri and Garfinkle's memory due to the outgoing state as \cite{Bieri:2013ada}
\begin{equation}
    -\bigg(E^{(3)}_{rr}(\infty)-\sum_{i=0,1}[E^{(3)}_{rr}(\oo)]_{[i]}\bigg)+8\pi\bigg(F-\sum_{i=0,1}[F]_{[i]}\bigg) \ .
\end{equation}
As we have discussed, in Bieri and Garfinkle's treatment, this is a first order result, and at this order $F$ does not include the energy from external gravitons. However, we will use this result with an amended $F$ which does include the energy due  to external gravitons. Here we have dropped contributions to the memory which are due to fields at $u<0$. These are not causally related to the outgoing state and are the contribution from the incoming state, which we are not interested in here (this will be encoded into the incoming Fock space rather than the outgoing Fock space). It is immediately clear that given $n'_g$ outgoing gravitons with momenta $p_{g'}$ the outgoing radiation due to massless particles per unit solid angle is given by 
\begin{equation}
    F=\sum_{g'=1}^{n'_g}\frac{\w_{g'}}{\gamma_{z\bar{z}}}\delta^2(z_{g'}-z_{y}) \ .
\end{equation}
From this we find
\begin{equation}
    [F]_{[0]}=\frac{1}{4\pi}\sum_{g'=1}^{n'_g}\int d^2 z'\gamma_{z'\bar{z}'}\frac{\w_{g'}}{\gamma_{z\bar{z}}}\delta^2(z_{g'}-z_{y}) =\frac{1}{4\pi}\sum_{g'=1}^{n'_g}\w_{g'}\ ,
\end{equation}
and
\begin{equation}
    [F]_{[1]}=\frac{3}{4\pi}\sum_{g'=1}^{n'_g}\int d^2z' \gamma_{z'\bar{z}'}  \ \hat{\by}'\cdot \hat{\by}\frac{\w_{g'}}{\gamma_{z\bar{z}}}\delta^2(z_{g'}-z') =\frac{3}{4\pi}\sum_{g'=1}^{n'_g}\bp_{g'}\cdot \hat{\by} \ .
\end{equation}
Since 
\begin{equation}
    -8\w_{g'}-6p_{g'}\cdot \hat{y}=-8\pi(\frac{1}{4\pi}\w_{g'}+\frac{3}{4\pi}\bp_{g'}\cdot \hat{\by}) \ ,
\end{equation}
We have shown 
\begin{equation}
    \sum_{g'=1}^{n'_g}(-8\w_{g'}-6p_{g'}\cdot \hat{y}+\frac{8\pi\w_{g'}}{\gamma_{z\bar{z}}}\delta^2(z_{g'}-z_y))=8\pi\bigg(F-\sum_{i=0,1}[F]_{[i]}\bigg) \ .
\end{equation}
Next we wish to calculate $E^{(3)}_{rr}(\oo)$ due to a massive particle with momentum $p$. We have defined $E_{rr}$  as 
\begin{equation}
    E_{rr}=C_{rtrt} \ .
\end{equation}
Here $C_{rtrt}$ is the Weyl tensor which is given in terms of the Riemann tensor by
\begin{equation}
    C_{abcd} = R_{abcd}-\frac{1}{2}(g_{ac}S_{db}-g_{ad}S_{cb}-g_{bc}S_{da}+g_{bd}S_{ca}) \ , 
\end{equation}
where
\begin{equation}
    S_{ab}=R_{ab}-\frac{1}{6}Rg_{ab} \ .
\end{equation}
Then  $E^{(3)}_{rr}(u)$ is the leading order term in a large $r$ expansion at constant retarded time $u$. In the linearized theory at first order in perturbation theory we have $S_{ab}=0$ far away from the scattering process, since the gravitational wave energy only contributes to the stress-energy at second order in perturbation theory. Thus, at the field point of interest we have $R_{abcd}=C_{abcd}$. Then we find 
\begin{equation}
    C_{ruru}=R_{ruru}=-\frac{1}{2}\partial^2_r \delta g_{uu}\ .
\end{equation}
In Bondi gauge, we have 
\begin{equation}
    -\frac{1}{2}\partial^2_r \delta g_{uu}=-\frac{1}{2}\partial^2_r\frac{2m_B}{r}= -\frac{2m_B}{r^3} \ , \qquad m_B=\frac{m^4}{(E-\bp\cdot\hat{\by})^3} \ ,
\end{equation}
where $m_B$ is the Bondi mass. Note that as noted by Bieri and Garfinkle in \cite{Bieri:2013ada} the Weyl tensor is gauge invariant at first order in perturbation theory since it vanishes in Minkowski spacetime. We thus find
\begin{equation}
    E_{rr}^{(3)}=\sum^{n'_s}_{s'=1}\frac{2m^4}{(p_{s'}\cdot \hat{y})^3} \ .
\end{equation}
This gives
\begin{equation}
    [E^{(3)}_{rr}]_{[0]}=\frac{1}{4\pi}\sum_{s'=1}^{n'_s}\int d^2 z\gamma_{z\bar{z}}\frac{2m^4}{(p_{s'}\cdot \hat{y})^3} =-\sum_{s'=1}^{n'_s}2\w_{s'}\ ,
\end{equation}
and
\begin{equation}
    [E^{(3)}_{rr}]_{[1]}=\frac{3}{4\pi}\sum_{s'=1}^{n'_s}\int d^2 z'\gamma_{z'\bar{z}'}\ \hat{\by}'\cdot \hat{\by}\,\frac{2m^4}{(p_{s'}\cdot \hat{y}')^3} =-6\sum_{s'=1}^{n'_s}\bp_{s'}\cdot \hat{\by} \ .
\end{equation}
Combining everything, we have shown
\begin{equation}
    \lambda(\hat{y})=-\bigg(E^{(3)}_{rr}(\infty)-\sum_{i=0,1}[E^{(3)}_{rr}(\oo)]_{[i]}\bigg)+8\pi\bigg(F-\sum_{i=0,1}[F]_{[i]}\bigg) \ .
\end{equation}
This is what we required, so we are done.

\acknowledgments
We thank Luke Lippstreu, Kartik Prabhu, Lydia Bieri, David Garfinkle, Ian Moult, and Sabrina Pasterski for useful discussions. We acknowledge that Claude was used to help check and prepare the latest version of this paper.
The author is supported by the DOE Early Career Award DE-SC0025581, the Sloan Foundation, and the Simons Collaboration on Confinement and QCD Strings.



\bibliographystyle{JHEP}
\bibliography{biblio.bib}






\end{document}